\documentclass[%
 reprint,
superscriptaddress,
 amsmath,amssymb,
 aps, prl,
floatfix,
]{revtex4-1}

\usepackage{graphicx} 
\usepackage{dcolumn} 
\usepackage{sidecap}
\usepackage{float}
\usepackage{bm} 
\usepackage[svgnames]{xcolor}
\usepackage[bookmarks=false,linkcolor=blue,urlcolor=blue,colorlinks,citecolor=blue]{hyperref}
\begin{document}

\title{Phase Asymmetry of Andreev Spectra From Cooper-Pair Momentum} 

\author{Abhishek Banerjee}
\affiliation{Center for Quantum Devices, Niels Bohr Institute,
University of Copenhagen,
Universitetsparken 5, 2100 Copenhagen, Denmark
}

\author{Max Geier}
\affiliation{Center for Quantum Devices, Niels Bohr Institute,
University of Copenhagen,
Universitetsparken 5, 2100 Copenhagen, Denmark
}

\author{Md Ahnaf Rahman}
\affiliation{Center for Quantum Devices, Niels Bohr Institute,
University of Copenhagen,
Universitetsparken 5, 2100 Copenhagen, Denmark
}

\author{Candice Thomas}
\affiliation{Department of Physics and Astronomy, and Birck Nanotechnology Center,
Purdue University, West Lafayette, Indiana 47907 USA
}

\author{Tian Wang}
\affiliation{Department of Physics and Astronomy, and Birck Nanotechnology Center,
Purdue University, West Lafayette, Indiana 47907 USA
}

\author{Michael J.~Manfra}
\affiliation{Department of Physics and Astronomy, and Birck Nanotechnology Center,
Purdue University, West Lafayette, Indiana 47907 USA
}

\affiliation{School of Materials Engineering, and School of Electrical and Computer Engineering, Purdue University, West Lafayette, Indiana 47907 USA}

\author{Karsten Flensberg}
\affiliation{Center for Quantum Devices, Niels Bohr Institute,
University of Copenhagen,
Universitetsparken 5, 2100 Copenhagen, Denmark
}

\author{Charles~M.~Marcus}
\affiliation{Center for Quantum Devices, Niels Bohr Institute,
University of Copenhagen,
Universitetsparken 5, 2100 Copenhagen, Denmark
} 

\date{\today} 

\begin{abstract}
In analogy to conventional semiconductor diodes, the Josephson diode exhibits superconducting properties that are asymmetric in applied bias. The effect has been investigated in number of systems recently, and requires a combination of broken time-reversal and inversion symmetries. We demonstrate a dual of the usual Josephson diode effect, a nonreciprocal response of Andreev bound states to a superconducting phase difference across the normal region of a superconductor-normal-superconductor Josephson junction, fabricated using an epitaxial InAs/Al heterostructure. Phase asymmetry of the subgap Andreev spectrum is absent in the absence of in-plane magnetic field and reaches a maximum at 0.15~T applied in the plane of the junction transverse to the current direction. We interpret the phase diode effect in this system as resulting from finite-momentum Cooper pairing  due to orbital coupling to the in-plane magnetic field, without invoking Zeeman or spin-orbit coupling.

\end{abstract}

\maketitle

Nonreciprocal effects in superconducting systems have attracted significant recent interest. A primary motivation is to engineer a superconducting diode, a circuit element that supports dissipationless flow of current in one direction but is resistive in the opposite~\cite{Hu2007Diode}. Nonreciprocity of supercurrent flow has been observed in thin film superconductors~\cite{Ryohei2017,qin2017superconductivity,yasuda2019nonreciprocal,itahashi2020nonreciprocal,Ando2020,bauriedl2021supercurrent,shin2021magnetic,diez2021magnetic,lin2021zero,sundaresh2022supercurrent} and Josephson junction devices~\cite{Avignon2008,pal2021josephson,wu2022field,baumgartner2022supercurrent,costa2022sign}. In general, nonreciprocal superconductivity requires the breaking of time-reversal and inversion symmetries~\cite{Martin2009,Brunetti2013Anomalous,Yokoyama2014anomalous,Chen2018,WakatsukiNonreciprocal2018,HoshinoNonreciprocal2018,pal2019quantized, he2022phenomenological,Bergeret2022supercurrent,yuan2022supercurrent,davydova2022universal,DaidoIntrinsic2022}, also prerequisites for various topological superconducting states. Nonreciprocal effects may therefore also serve as a marker for unconventional superconducting order~\cite{Brunetti2013Anomalous,Chen2018,zinkl2021symmetry,diez2021magnetic,lin2021zero}.  

Several proposals have been put forward to explain nonreciprocal superconducting transport. In uniform noncentrosymmetric materials, breaking of time-reversal symmetry with a magnetic field generates a nonreciprocal supercurrent~\cite{HoshinoNonreciprocal2018,Ando2020,DaidoIntrinsic2022,yuan2022supercurrent,he2022phenomenological,Bergeret2022supercurrent}. In Josephson junctions, various mechanisms may lead to a Josephson diode effect including a combination of spin-orbit coupling and external magnetic field~\cite{Martin2009,Brunetti2013Anomalous,Yokoyama2014anomalous,Chen2018}, ferromagnetism of the barrier material~\cite{pal2019quantized,Bergeret2022rectification}, and finite-momentum Cooper pairing ~\cite{zinkl2021symmetry,pal2021josephson,DavydovaSci2022}.

The behavior of Andreev bound states (ABSs) carrying nonreciprocal supercurrents across Josephson junctions has been predicted in the form of an asymmetric energy-phase relationship where $E(\phi_0-\phi)\neq E(\phi_0 + \phi)$ for any value of $\phi_0$~\cite{Yokoyama2014anomalous,DavydovaSci2022}. This is different from the $\phi_0$ Josephson effect where the ABS spectrum is phase-shifted by $\phi_0$ but still remains symmetric about $\phi=\phi_0$~\cite{BuzdinPRL2008,DolciniPRB2015,Nesterov2016Anomalous,szombati2016josephson}. Signatures of phase-asymmetric ABSs have been observed in semiconductor Josephson junctions where this behavior was attributed to spin-orbit coupling~\cite{van2017microwave,Tosi2019spin}.

Here, we use a combination of local and nonlocal tunneling spectroscopy to experimentally investigate the ABS spectrum of planar InAs/Al Josephson junctions  as a function of superconducting phase difference, $\phi$, across the junction, and magnetic field, $B_\parallel$, applied in the plane of the sample, parallel to the supercondutor-normal (SN) interfaces. At zero applied magnetic field, the ABS spectrum is symmetric under phase-inversion. A small magnetic field, $B_\parallel \simeq$~60~mT results in an ABS spectrum with a pronounced phase-asymmetry that becomes maximal at $B_\parallel \simeq$~0.15~T. With three-terminal conductance spectroscopy, we show that these phase-asymmetric ABSs are present throughout the bulk of the Josephson junction, as opposed to only the device ends.  

\begin{figure}[!bt]
\includegraphics[width=0.45\textwidth]{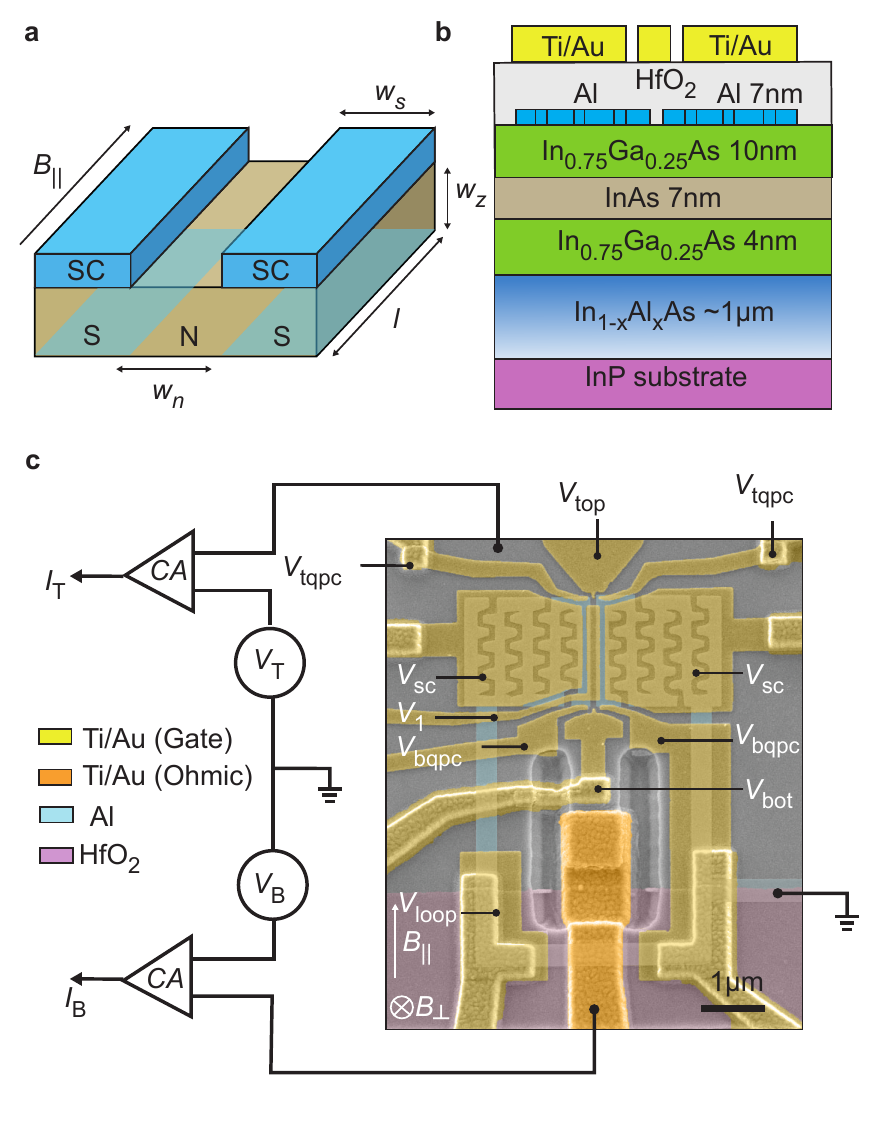}
\caption{\label{fig01} {\bf Device schematic and micrograph}  
(a) Schematic of a planar Josephson junction device consisting of two superconducting leads (blue) of width $w_s= 1.8 \,\mu$m in epitaxial contact with the underlying semiconductor (brown). The normal region between the two leads is of width $w_n = 100$~nm, length $l= 1.6 \,\mu$m. The 2D electron wave-function extends over a nominal thickness of $w_z\sim 20$~nm that includes the width of the InAs quantum well and the two insulating In$_{0.75}$Ga$_{0.25}$As barriers. (b) Cross-section of the device depicted schematically showing various layers comprising the heterostructure stack. Al (7~nm) is epitaxially grown on an In$_{0.75}$Ga$_{0.25}$As (10~nm)/InAs (7~nm)/In$_{0.75}$Ga$_{0.25}$As (4~nm) quantum well. Meandering perforations in the Al leads help with hardening of the superconducting gap. (c) ~False-colored electron micrograph of a planar Josephson junction device measured in a three-terminal configuration. DC biases, $V_{\rm T}$ and $V_{\rm B}$, are applied to the top and bottom ohmics through current amplifiers connected to the respective terminals. The superconducting loop is grounded. Gates $V_{\rm top(bot)}$ and $V_{\rm t(b)qpc}$ create an electrostatic constriction at the top (bottom) end for tunneling spectroscopy. $V_{\rm sc}$ controls density under the superconducting leads. $V_{1}$ controls density in the junction. An out-of-plane magnetic field $B_\perp$ threads magnetic flux through the superconducting loop for phase biasing. In-plane magnetic field $B_\parallel$ is applied in the plane of the device parallel to the S-N interfaces.}
\end{figure}

Based on numerical simulations, we interpret the observed phase-asymmetry of the spectrum as a signature of ABSs that carry nonreciprocal supercurrents, which we denote {\it nonreciprocal} ABSs. Although InAs has strong spin-orbit coupling and a large g-factor, we find they are insufficient to explain the strong nonreciprocal effects we observe at relatively low magnetic fields. Instead, we find consistency with a mechanism where the orbital effect of $B_\parallel$ creates a superconducting state in the junction with finite-momentum Cooper pairing, similar to a mechanism proposed in recent works~\cite{pal2021josephson,DavydovaSci2022}. Although our device geometry prevents us from performing supercurrent measurements, with model parameters that fit our Andreev bound state spectrum, we estimate a supercurrent diode efficiency of up to $\sim$10\%. 

The planar SNS devices we study are illustrated schematically in Fig.~\ref{fig01}(a). The S regions are created by inducing superconductivity into a semiconducting material by a parent superconductor (SC) via proximity effect. 
Devices are fabricated on InAs/Al two-dimensional heterostructures shown in Fig.~\ref{fig01}(b), where Al plays the role of the parent superconductor (SC), and InAs is the semiconducting layer that serves as both the proximitzed superconductor (S) where Al is present, and the normal (N) region of the junction where Al has been removed. The active semiconductor region, consisting of an InAs quantum well confined by  In$_{0.75}$Ga$_{0.25}$As barriers, has a thickness of $w_z\sim$~20~nm. 
As we indicate later, this thickness combined with behavior of S and SC as distinct superconducting elements, allows the in-plane magnetic field to have a strong orbital effect generating finite-momentum Cooper pairing in S~\cite{volovik2007quantum,zhu2021discovery,hart2017controlled}.

Figure~\ref{fig01}(c) shows an electron micrograph of one of the devices, along with a schematic electrical circuit. The device geometry, fabrication method, and measurement setup has been discussed in detail in previous works~\cite{Banerjee2022topological,banerjee2022local}. Briefly, the Josephson junction device has a superconducting loop that connects the two superconducting leads, allowing phase-biasing by the application of a small (0.1~mT scale) out-of-plane magnetic field, $B_\perp$. Spectroscopy is performed by quantum point contacts (QPC) at both ends of the junction, formed by split gates that are controlled by voltages $V_{\rm tqpc}$ and $V_{\rm bqpc}$ at the top and bottom ends. The carrier density in the normal barrier (width $w_n=100$~nm, length $l=1.6\,\mu$m) is controlled by gate voltage $V_1$, while the carrier density in the semiconductor below the superconducting leads is controlled by gate voltage $V_{\rm sc}$. Additional gate voltages $V_{\rm top}$ and $V_{\rm bot}$ control densities in the normal regions outside the QPCs, and are typically fixed at $\sim 100$~mV. We discuss results from Device 1 and Device 2 in the main text, and results from Device 3 are provided in the Supplementary Material. 

\begin{figure*}[t]
\includegraphics[width=1.0\textwidth]{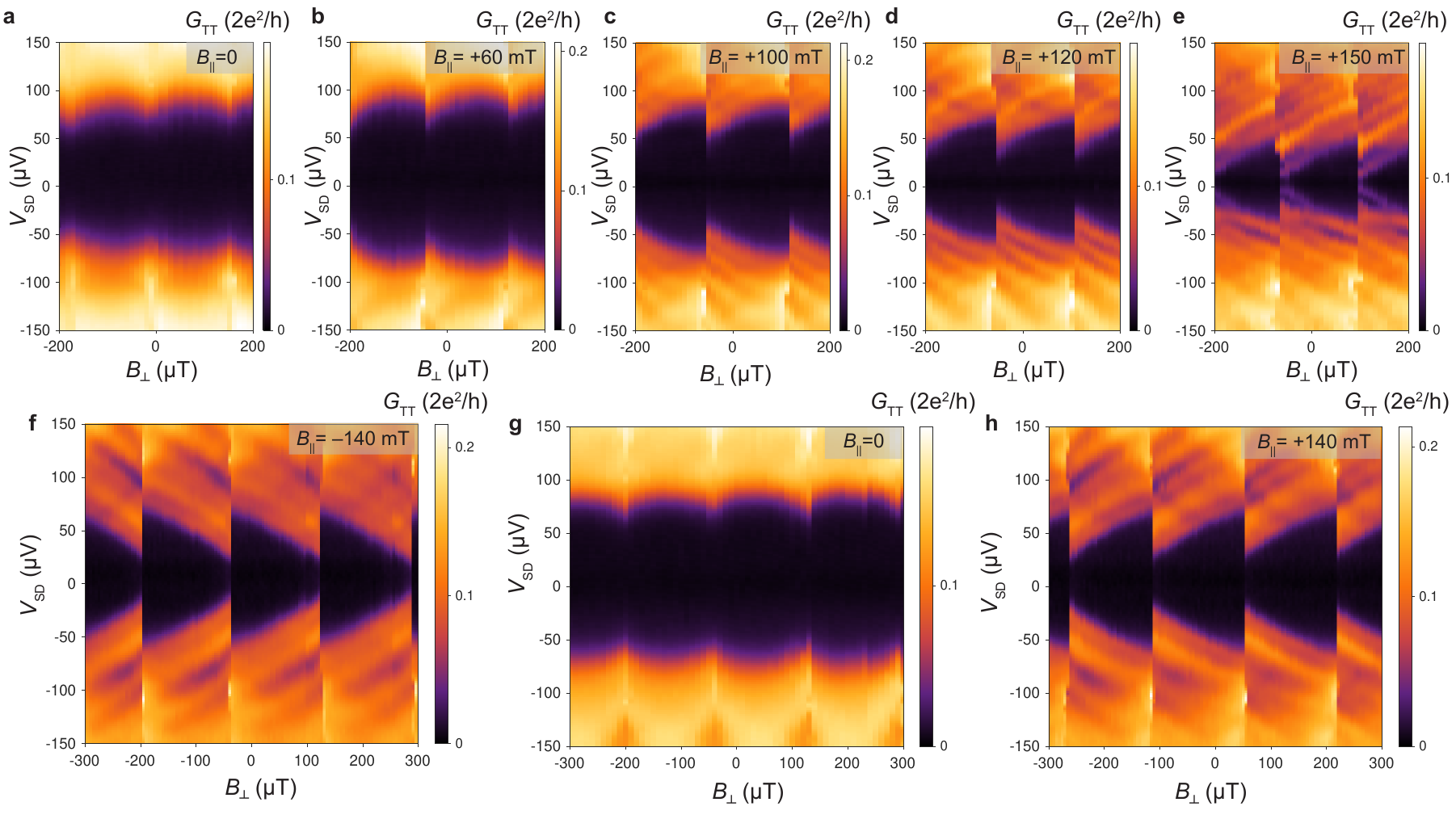}
\caption{\label{fig02} 
{\bf Nonreciprocal Andreev bound state spectrum.} (a)--(e) Differential conductance measured in Device 1 as a function of $B_\perp$ at different value of $B_\parallel$. At $B_\parallel=$0, the Andreev bound state spectrum is phase-symmetric within each flux lobe. For non-zero $B_\parallel$, the spectrum acquires an asymmetry within each flux lobe. The asymmetry is strongest at $B_\parallel=0.15$~T. (f)-(h) The sense of asymmetry within each flux lobe is opposite for $B_\parallel=+0.14$~T and (h) $B_\parallel=-0.14$~T. There is no asymmetry at (g) $B_\parallel=0$. }
\end{figure*}

We first focus on Device 1, where we performed tunneling spectroscopy only using the top QPC. At zero in-plane magnetic field, we observe a periodic modulation of the superconducting gap with period $B_\perp \simeq 0.16~$~mT corresponding to one flux quantum $\Phi_0=h/2e$ through the superconducting loop of area $\sim15~\mu$m$^2$. Within each flux lobe, the gap modulation appears symmmetric around maxima and minima, as shown in Figs.~\ref{fig02}(a, g). This symmetry is broken by the application of $B_\parallel$. Already at $B_\parallel \simeq$~0.06~T [Fig.~\ref{fig02}(b)], the flux lobes are lopsided, with a smaller gap measured at the left end of the flux lobe compared to the right. Further increasing $B_\parallel$ increased the lopsidedness, while the overall superconducting gap was also reduced. At $B_\parallel=0.15$~T [Fig.~\ref{fig02}(e)], a discrete set of differential conductance peaks emerge, with an arrow-like structure pointing toward the left end of the flux lobe. Reversing the direction of the in-plane magnetic field reverses the direction of the arrow, as shown in the comparison of datasets obtained at $B_\parallel=-0.14$~T [Fig.~\ref{fig02}(f)]  and $B_\parallel=+0.14$~T [Fig.~\ref{fig02}(h)]. Flux lobes are separated by sharp discontinuities in differential conductance at finite magnetic field. At $B_\parallel=0$, the discontinuity is absent or at least not noticeable. The discontinuities become larger as the spectrum becomes more asymmetric.
 
\begin{figure}[t]
\includegraphics[width=0.45\textwidth]{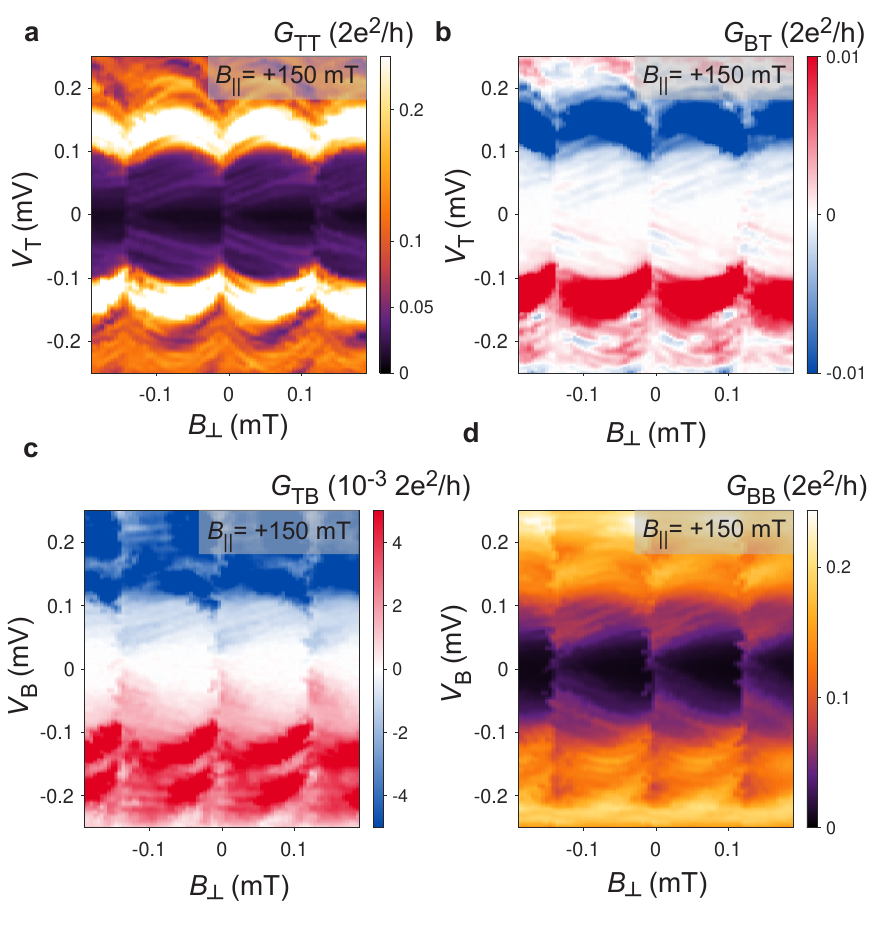}
\caption{\label{fig03} {\bf Nonreciprocal differential conductance matrix.} Local conductances (a) $G_{\rm TT}$ and (d) $G_{\rm BB}$ , and nonlocal conductances (b) $G_{\rm BT}$ and (c) $G_{\rm TB}$, measured as a function of source-drain biases $V_{\rm T,B}$ and out-of-plane magnetic field $B_\perp$ at $B_\parallel=+0.14$~T, in Device 2. Both local and nonlocal conductances show phase-asymmetric Andreev bound states that have a smaller gap at the left end of a flux lobe compared to the right end. Phase symmetric Andreev bound states are also visible in both local and nonlocal conductances at high source drain biases $|V_{\rm T,B}| \geq 100~\mu$eV.}
\end{figure}

Three-terminal conductance spectroscopy can be used to further probe nonreciprocity of the ABS spectrum. Measurements of the 2$\times$2 matrix of differential conductance for Device 2 are shown in Fig.~\ref{fig03}. At $B_\parallel=0.15$~T, we observe strongly phase-asymmetric local conductances, $G_{\rm TT}$ and $G_{\rm BB}$ at the two device ends, as well as asymmetric nonlocal conductances, $G_{\rm TB}$ and $G_{\rm BT}$.  Compared to the spectrum measured in Device 1, we note that the phase-asymmetric differential conductance peaks have more discrete structure, representing individual energy levels that can be recognized in all 4 conductance matrix elements. Additionally, we observe a set of phase-symmetric conductance features that exist at high source-drain bias voltages $V_{\rm T/B}\simeq 0.1$~mV, close to the superconducting gap edge. At zero magnetic field, phase-symmetry of the differential conductance matrix is restored (see Figs.~\ref{figS16},~\ref{figS17} in the Supplementary Material).

Nonlocal spectra indicate that phase-asymmetric ABSs  extend throughout the bulk of the junction, as opposed to localized features associated only with the device ends~\cite{DanonNonlocal,Maiani2022}. Extended ABSs mediate nonlocal transport  between the two device ends, leading to a finite phase-asymmetric nonlocal conductance signal. This is crucial as localized ABSs may have phase-asymmetric behavior that is different at the two ends and in the bulk, and has been shown to be associated with Josephson vortex physics~\cite{Banerjee2022vortex}.

\begin{figure}[t]
\includegraphics[width=0.45\textwidth]{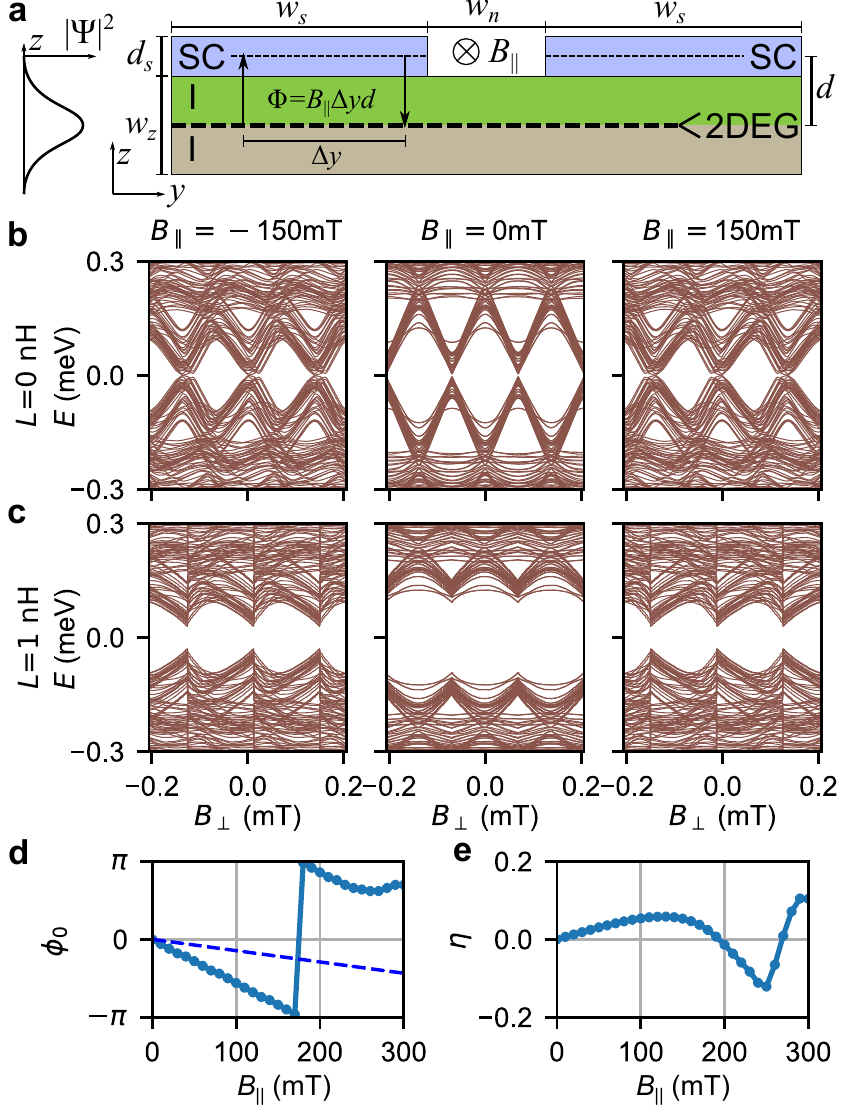}
\caption{\label{fig04} \textbf{Model and numerical results.}
(a) Sketch of the Josephson junction cross section formed by two superconducting films of width $w_s = 400$~nm and separation $w_n = 100$~nm deposited on top of a 2DEG with wavefunction profile $|\Psi|^2$ (sketch). The mean separation between the 2DEG and the superconductor is $d \sim 15$~nm.
Cooper pairs in the proxmitized 2DEG acquire a finite momentum $q = \pi B_\parallel d / \Phi_0$ from the flux $\Phi = B_\parallel d \Delta y$ enclosed by electron hopping between the layers at different positions $\Delta y$. 
(b), (c) Calculated spectrum in the two-dimensional geometry with in-plane field $B_x = -150,\  0,\ \text{and}\ 150$~mT (from left to right) with loop inductance $L=0$ (b) and $L = 1$~nH (c). 
(d) Phase difference $\phi_0$ minimizing the junction potential energy (solid line) and phase bias $- 2 q w_n$ imposed by the Cooper pair momentum $q$ (dashed line). (e) Diode efficiency $\eta = (I_C^+ - |I_C^-|)/(I_C^+ + |I_C^-|)$.
}
\end{figure}

We turn to numerical simulation to better understand the observed nonreciprocity. 
The InAs quantum well is modelled as a two-dimensional electron gas (2DEG) with effective mass $m^* = 0.026\, m_e$ \cite{VurgaftmanJAP2001}, chemical potential $\mu=1.5$~meV \cite{Banerjee2022topological, Banerjee2022vortex}, spin-orbit coupling $\alpha = 15$~meV~nm\cite{MayerACS2020, BaumgartnerNNano2022}, and $g$-factor $g=-10$ \cite{SmithPRB1987, MayerACS2020, BaumgartnerNNano2022} in proximity to an Al superconducting layer with superconducting gap $|\Delta| = 0.2$~meV \cite{CohenPR1968}. Accounting for the thickness of the quantum well $w_z = 20$~nm and superconductor $d_s = 7$~nm, the 2DEG is located $d = (w_z + d_s)/2 \sim 15$~nm below the superconducting layer [Fig.~\ref{fig04}(a)].
Electrons hopping between the two layers acquire a geometric phase shift from the orbital coupling to the in-plane magnetic field $B_\parallel$. As a consequence, Cooper pairs in the proximitized 2DEG acquire a finite momentum perpendicular to the applied field. In the gauge $\vec{A} = - B_\parallel z \hat{e}_y$, the orbital coupling yields a phase gradient $\phi \to \phi + 2 q y$ of the proximity-induced pairing $\Delta e^{i \phi}$ \cite{Pientka} where $q = \pi B_\parallel d / \Phi_0 $ is the Cooper pair momentum \cite{DavydovaSci2022} (see Supplementary Material for details).
The Cooper pair momentum yields a Doppler shift of forward and backward moving quasiparticles, which results in a nonreciprocal quasiparticle spectrum and current-phase relation \cite{DavydovaSci2022}. 

This is captured by our numerical spectrum shown in Fig.~\ref{fig04}(b). At $B_\parallel = 0$, the spectrum is symmetric. At $B_\parallel = \pm 0.15$~T, the spectrum becomes nonreciprocal, with opposite asymmetries at opposite $B_\parallel$, consistent with Onsager micro-reversibility. We further include the effect of the inductance $L \sim 1$~nH of the superconducting loop on the observed spectrum. The inductance creates a competition between the junction potential energy $E_\text{J}(\phi)$ and kinetic energy $L I^2(\phi)/2$. This is evaluated numerically (see Supplementary Material) and results shown in Fig.~\ref{fig04}(c). Qualitatively, the inductance has two effects: i) it reduces the width of the flux lobe from $\Delta \phi=2\pi$ because a portion of the applied phase-bias now drops across the superconducting loop $\simeq 2 \pi IL/\Phi_0$. As a consequence, sharp phase-jumps are observed at the ends of each flux lobe. ii) the center of the flux lobe is shifted away from $\phi=0$ to $\phi=\phi_0$ where $\phi_0$ is the phase that minimizes the junction potential energy $E_\text{J}(\phi)$, shown in Fig.~\ref{fig04}(d) as a function of $B_\parallel$ (see Supplementary Material for details).  We note that the inductance does not create any phase-asymmetry, as seen from the data and numerics at $B_\parallel=0$, but does help to  amplify its visibility at finite $B_\parallel$.

The nonreciprocal spectrum can be further characterized by the asymmetry of the zero-temperature critical currents $I_C^\pm = \frac{2e}{\hbar} \max (\pm \partial E_\text{J}(\phi)/\partial \phi)$ in terms of the diode efficiency $\eta = (I_C^+ - |I_C^-|)/(I_C^+ + |I_C^-|)$. These are the calculated junction critical currents for a current-biased geometry excluding the loop. As shown in Fig.~\ref{fig04}(e), $\eta \simeq$~8\% at $B_\parallel \simeq 150$~mT, roughly where the ABS spectrum is most phase-asymmetric. For material parameters used here, nonreciprocity arising from finite-momentum Cooper pairing dominates contribution arising from the interplay between spin-orbit and Zeeman couplings, the more commonly explored mechanism for nonreciprocal behavior~\cite{BuzdinPRL2008, YokoyamaPRB2014, DolciniPRB2015, BaumgartnerNNano2022}. Nonreciprocal ABSs can also be obtained without assuming finite-momentum Cooper pairing, but requires inflated values of $g$-factor ($|g|\sim 60$) and spin-orbit coupling ($\alpha \sim 90$~meV~nm), which are unreasonable values for this material (see Supplementary Material).

Good agreement between theory and experiment supports our interpretation of finite-momentum Cooper pairing arising from the orbital coupling of the in-plane magnetic field as the dominant mechanism for the experimentally observed nonreciprocity. This mechanism does not rely on spin-orbit or Zeeman couplings, and indicates that other proximity-effect based superconducting platforms such as graphene \cite{LeeNatComm2015}, Ge/SiGe \cite{ScappucciNatRev2021}, or GaAs \cite{WanNatComm2015} heterostructures, where these couplings are small or nonexistent, may also host nonreciprocal superconductivity.  Recent studies in similar devices \cite{Banerjee2022topological, Sundaresh2022} have also indicated the relevance of orbital physics in the search for topological superconductivity \cite{HellFlensberg, Pientka, lesser2021phase,lesser2021three}. The connection between finite-momentum Cooper pairing, nonreciprocal superconductivity, and topological superconductivity  warrents further study \cite{QuNatComm2013, MeloSciPost2019}.

We thank Geoff Gardner and Sergei Gronin for contributions to materials growth, and  Asbj{\o}rn Drachmann for assistance with fabrication. We acknowledge a research grant (Project 43951) from VILLUM FONDEN. MG acknowledges support by the European Research Council (ERC) under the European Union’s Horizon 2020 research and innovation program under grant agreement No.~856526, and from the Deutsche Forschungsgemeinschaft (DFG) project grant 277101999 within the CRC network TR 183 (subproject C01), and from the Danish National Research Foundation, the Danish Council for Independent Research $\vert$ Natural Sciences.

\bibliography{Nonreciprocal_bib.bib}

\clearpage

\onecolumngrid
\appendix
\begin{center}
{\bf Supplementary Material for \\Phase Asymmetry of Andreev Spectra From Cooper-Pair Momentum}
\end{center}

\setcounter{equation}{0}
\renewcommand{\theequation}{S\arabic{equation}}
\setcounter{figure}{0}
\renewcommand{\thefigure}{S\arabic{figure}}
\setcounter{section}{0}
\renewcommand{\thesection}{S\Roman{section}}
\setcounter{table}{0}
\renewcommand{\thetable}{S\Roman{table}}

\vspace{10 pt}

{\bf Wafer structure:} The wafer structure used in this work consists of an InAs two-dimensional quantum well in epitaxial contact with Al, grown by molecular beam epitaxy. The wafer was grown on an insulating InP substrate and comprises a 100-nm-thick In$_{0.52}$Al$_{0.48}$As matched buffer, a 1$\mu$m thick step-graded buffer realized with alloy steps from In$_{0.52}$Al$_{0.48}$As to In$_{0.89}$Al$_{0.11}$As (20 steps, 50 nm/step), a 58 nm In$_{0.82}$Al$_{0.18}$As layer, a 4 nm In$_{0.75}$Ga$_{0.25}$As bottom barrier, a 7 nm InAs quantum well, a 10 nm In$_{0.75}$Ga$_{0.25}$As top barrier, two monolayers of GaAs and a 7 nm film of epitaxially grown Al. The top Al layer was grown without breaking the vacuum, in the same growth chamber.  Hall effect measurements performed in Hall bar devices of the same material, with Al etched away, indicated a peak electron mobility peak $\mu= 43,000$~ cm$^2$/Vs at a carrier density of $n=8 \times 10^{11}$~cm$^{-2}$, corresponding to a peak electron mean free path of  $l_e\sim$~600 nm suggesting that our devices are quasi-ballistic along the length $l \simeq 3 l_e$ and ballistic in the width direction $w_n \simeq l_e/6$. Transport characterization of an etched Al Hall bar revealed an upper critical field of 2.5~T, indicating that the Al layer loses superconductivity at a field much larger than the collapse of the induced superconducting gap at $\sim$~0.5~T.   

\vspace{10 pt}

{\bf Device fabrication:} Devices were fabricated with standard electron beam lithography techniques. Devices on the same chip were electrically isolated from each other using a two-step mesa etch process, first by removing Al with Al etchant Transene D, and then a standard III-V chemical wet etch using a solution comprising H$_2$O : C$_6$H$_8$O$_7$ : H$_3$PO$_4$ : H$_2$O$_2$ (220:55:3:3) to etch the mesa until a depth of $\sim$300~nm. This step also defined the U-shaped trench and the patch of mesa that eventually formed the submicron ohmic contact. In the next lithography step, the Al layer was selectively removed, leaving behind the Josephson junction with a flux loop. This step also removed Al from the internal ohmic mesa patch. A patterned layer of dielectric, comprising 15 nm thick HfO$_2$ grown at 90$^\circ$C using atomic layer deposition (ALD), was then deposited to galvanically isolate the ohmic Ti/Au layer from the rest of the device. This was followed by the deposition of Ti/Au layers (5~nm/300~nm) for the inner ohmic contact. Next, a global layer of 15 nm thick HfO$_2$ was deposited over the entire sample to serve as the gate dielectric. Gates were defined using electron beam lithography followed by e-beam evaporation of Ti/Au layers of thickness (5nm/20nm) for finer structures and (5nm/350nm) for the bonding pads.

\vspace{10 pt}

{\bf Transport measurements:} Electrical transport measurements were performed using an Oxford Triton 400 dilution refrigerator at a base temperature of $\sim$20 mK. The superconducting loop was grounded by connecting it to the fridge ground through $\sim$900$\,\Omega$. Top and bottom ohmic contacts were connected to low-impedance current-to-voltage converters through  $\sim$1.5--2.0~k$\Omega$ filter and cable resistances. The grounding pin of each current-to-voltage converter was biased through a combination of AC+DC voltages $V_{\rm t(b)}+V_{\rm T(B)}$, at the top and bottom ends respectively. The AC biases, $V_{\rm t}$ and $V_{\rm b}$, were generated by two lock-in amplifiers, with the same excitation amplitude (3~$\mu$V), but different frequencies, $f_{\rm t}=31.5$~Hz and $f_{\rm b}=77.5$~Hz respectively. The DC biases $V_{\rm T(B)}$ were generated using two low-noise DC voltage sources. The voltage output of each current-to-voltage converter was measured using two lock-in amplifiers operating at frequencies $f_{\rm t}$ and $f_{\rm b}$, requiring four lock-in amplifiers in total, one for each element of the 2$\times$2 conductance matrix.

\vspace{10 pt}

{\bf Phase biasing and magnetic field alignment:}  Magnetic field to the sample is applied using a three-axis ($B_x$, $B_y$, $B_z$)=(1T, 1T, 6T) vector magnet. The sample is oriented with respect to the vector magnet such that $B_x$ is nominally parallel to $B_\perp$ [Fig.~\ref{fig01}(a)]. 

\vspace{10 pt}

{\bf Phase-profile in superconducting leads:}
Within each Al superconducting lead, an in-plane magnetic field $B_\parallel$ (i.e. pointing in $x$ direction, see Fig.~\ref{fig04}(a) for the geometrical sketch) generates a Meissner current density flowing perpendicular to the applied field.
As justified below, for this geometry we may assume that the Meissner current density is homogeneous along the magnetic field direction $x$ and neglect the magnetic self-field generated by the Meissner currents. In this case, the Meissner current density can be calculated from a Poisson equation for the superconducting phase profile analogously to Ref.~\cite{clem2010}. The resulting current density within the rectangular superconducting leads has elliptical contours of equal absolute current density with increasing current density from center to edge \cite{clem2010}. Our assumptions are justified as the self-field generated by the Meissner currents $B_{\parallel, \text{self}} \sim \frac{\text{min}(d_s, w_s)}{2\pi \Lambda} B_\parallel \sim 0.016 B_\parallel$ \cite{Banerjee2022vortex}, where $\Lambda = 2 \lambda_{L, \text{eff}}^2 / l = 73$~nm is the Pearl length for this geometry (homogeneous supercurrent density along $\hat{x}$ parallel to the applied field $(B_\parallel, 0, 0)^{\rm T}$) and $\lambda_{L, \text{eff}} = \lambda_{L, \text{Al}} \sqrt{\xi_{\text{Al}}/d_s} = 240$~nm is the effective London penetration depth of the Al thin film, see Section 3.11.4 of Ref.~\onlinecite{tinkham_introduction_2004} for reference on $\lambda_{L, \text{eff}}$ for thin films. 
Due to the large aspect ratio $w_s / d_s$ of the thin Al film, variations of the superconducting phase beyond the phase gradient discussed below are negligible \cite{clem2010, Banerjee2022vortex}.

In a gauge $\vec{A} = (0, -B_x z, 0)$ in which the vector potential vanishes within the plane of the two-dimensional electron gas at $z=0$, the superconducting order parameter in the Al film needs to have a phase gradient such that there is no net supercurrent in the Al film. This follows as the center of the Al film is located a distance $z = d$ above the center of the two-dimensional electron gas, such that the condition of zero net supercurrent reads
\begin{equation*}
    0 = j_y = - \frac{1}{\mu_0 \lambda_{L, \text{eff}}^2} \left( A_y + \frac{\Phi_0}{2\pi} \partial_y \phi \right)
\end{equation*}
where $A_y = - B_x d$ is the vector potential in the center of the Al film and $\phi$ is the phase of the superconducting order parameter in the Al film. This condition requires the phase gradient $ \partial_y \phi = \frac{2\pi}{\Phi_0} B_x d$. 

\vspace{10 pt}

{\bf Estimate of the kinetic inductance:}
The kinetic inductance of the superconducting loop can be expressed as
\begin{equation*}
    L_\text{kin} = \mu_0 \lambda_{L, \text{eff}}^2 \frac{U_s}{S_s} \sim 1\,\text{nH},
\end{equation*}
where $\mu_0 = 4\pi \times 10^{-7}\, \text{H}/\text{m}$, $\lambda_{L, \text{eff}} = \lambda_{L, \text{Al}} \sqrt{\xi_{\text{Al}}/d_s} = 240$~nm is the effective London penetration depth of the Al thin film (see Section 3.11.4 of Ref.~\cite{tinkham_introduction_2004}), $U_s \sim 18~\mu$m the circumference of the superconducting loop including the meander structure, and $S_s \sim 140~\text{nm}^2 $ is the cross-section of the superconducting stripe forming the loop \cite{MeserveyJAP1969}. This estimate agrees with the estimate in Ref.~\onlinecite{Banerjee2022topological} in which the kinetic inductance has been estimated using a measurement of the normal state resistivity. The two approaches are equivalent when applying the Drude model for the normal-state resistivity $\rho = m^* / (n_e e^2 \tau)$ using a mean free time $\tau = d_s / v_F$ limited by boundary scattering.

\vspace{10 pt}

{\bf Theory details:}  
We employ a tight-binding discretization of the spin-orbit coupled two-dimensional electron gas (2DEG) in the InAs quantum well including the coupling to the proximitized Al superconductors in terms of a phenomenological order parameter $\Delta$, and metallic leads to compute the spectrum and supercurrent from exact diagonalization. We use the software package \textit{kwant} \cite{groth_kwant_2014} for the Hamiltonian construction.
The discretized tight-binding Hamiltonian is
\begin{align}
\label{eq:S_H_N}
    H & =\sum_{\sigma,\sigma^{\prime}=\uparrow,\downarrow}\sum_{\vec{r}}\left(\left(\frac{\hbar^{2}}{m^{*}a^{2}}-\mu\right)\delta_{\sigma,\sigma^{\prime}}+\sum_{j=x,y,z}\frac{1}{2}g\mu_{B}B_{j}\sigma_{\sigma,\sigma^{\prime}}^{j}\right)\hat{c}_{\vec{r},\sigma}^{\dagger}\hat{c}_{\vec{r},\sigma^{\prime}}\\
	\ & +\sum_{\sigma,\sigma^{\prime}=\uparrow,\downarrow}\sum_{\vec{r}}\left(\left(\frac{\hbar^{2}}{4m^{*}a^{2}}\delta_{\sigma,\sigma^{\prime}}+\frac{i\alpha}{2}\sigma_{\sigma,\sigma^{\prime}}^{y}\right)\hat{c}_{\vec{r}+\vec{a}_{x},\sigma}^{\dagger}\hat{c}_{\vec{r},\sigma}+\left(\frac{\hbar^{2}}{4m^{*}a^{2}}\delta_{\sigma,\sigma^{\prime}}-\frac{i\alpha}{2}\sigma_{\sigma,\sigma^{\prime}}^{x}\right)\hat{c}_{\vec{r}+\vec{a}_{y},\sigma}^{\dagger}\hat{c}_{\vec{r},\sigma}+\text{h.c.}\right) \nonumber
\end{align}
with the effective mass $m^* = 0.026\,m_e$, lattice spacing $a=10$~nm, chemical potential $\mu = 1.45$~meV, $g$-factor $g=-10$, Bohr magneton $\mu_B$, Rashba spin-orbit coupling strength $\alpha = 15$~meV~nm, Kronecker-delta $\delta_{i,j}$ and Pauli matrices $\sigma^{j}, \ j=x,y,z$ in spin space. 
We include the orbital effect of the out-of-plane magnetic field $B_\perp$ via the substitution of the hopping terms
$$
\hat{c}_{\vec{r}+\vec{a}_{x/y},\sigma}^{\dagger}\hat{c}_{\vec{r},\sigma}\to e^{-i\frac{e}{\hbar}\int_{\vec{r}}^{\vec{r}+\vec{a}_{x/y}}\vec{A}d\vec{r}}\hat{c}_{\vec{r}+\vec{a}_{x/y},\sigma}^{\dagger}\hat{c}_{\vec{r},\sigma}.
$$
The proximity-induced pairing is included via a phenomenological s-wave, spin-singlet order parameter $\Delta(\vec{r})$ as \cite{McMillanPR1968, HellFlensberg}
\begin{equation}
\label{eq:S_H_S}
    H_\Delta = \sum_{\sigma,\sigma^{\prime}=\uparrow,\downarrow}\sum_{\vec{r}}\left(\Delta(\vec{r}) i\sigma_{\sigma, \sigma^\prime}^y \hat{c}_{\vec{r},\sigma}^{\dagger}\hat{c}_{\vec{r},\sigma^{\prime}}^\dagger +\text{h.c.} \right).
\end{equation}
The phase profile $\phi(\vec{r}) = \phi_{\text{L,R}} + \phi_\parallel(y) + \phi_\perp(x,y)$ of the superconducting order parameter $\Delta(\vec{r}) = |\Delta(\vec{r})|e^{i \phi(\vec{r})}$ includes a contribution from the flux through the loop $\phi_{\text{L,R}} = \pm \pi B_\perp A_\text{loop} / \Phi_0$ for the left / right superconducting lead, 
the phase gradient due to the in-plane magnetic field $\phi_\parallel(y) = 2\pi B_\parallel d y /\Phi_0$, and a phase texture $\phi_\perp(x,y)$ due to the out-of-plane magnetic field $B_\perp$ calculated following Ref.~\onlinecite{clem2010}. The latter contribution is small, $\phi_\perp(x,y) \ll 1$ because the out-of-plane magnetic field $B_\perp = \Phi_0/A_\text{loop}$ required to thread a flux quantum through the loop is small on the scale of the magnetic flux quantum through the junction area, $B_\perp (w_s + w_n) l / \Phi_0 \ll 1$. 
We approximate the magnitude of the phenomenological order parameter as $|\Delta| = 0.2$~meV in the regions that are covered by the superconducting banks, and zero otherwise. The value $|\Delta| = 0.2$~meV agrees with the parent gap of the Al superconductor.
The system is simulated in a rectangular region of length $l=1600$~nm and total width $w = w_n + 2 w_s$, where the proximity induced pairing is present only within the range $w_n/2 < |y| < w_n/2 + w_s$ with the width of the normal region $w_n = 100$~nm and the width of the superconducting leads $w_s = 400$~nm (unless stated otherwise, see Figs. \ref{figS11} to \ref{figS14} for calculations with $w_s = 200$~nm and $w_s = 600$~nm).

To compute the phase, $\phi_0$, that minimizes the ground state energy $E_J(\phi)$, zero-temperature supercurrents $I(\phi) = \frac{2e}{\hbar} \partial_\phi E_J(\phi)$ and diode efficiency $\eta$, we sum over the 400 negative eigenenergies with smallest absolute value $E_J(\phi) = \sum_n E_n(\phi)$. We verified that for this number the supercurrents and phase minimum $\phi_0$ are converged. 

To compute the superconducting phase bias $\phi$ as a function of perpendicular magnetic field including the flux jumps due to the inductance $L$ of the superconducting loop \cite{Banerjee2022topological}, we
require that the flux through the loop is screened by the supercurrent modulo $n \in \mathbb{Z}$ flux quanta,
\begin{equation}
    L I (\phi) + \frac{\Phi_0}{2 \pi} \phi - A_\text{loop} B_\perp + n \Phi_0 = 0,
\end{equation}
where the first term describes the phase drop along the superconducting loop, the second term describes the phase drop across the junction, and the third terms is the magnetic flux penetrating the loop.
This equation has multiple solutions. The physical system chooses to minimize its total energy 
\begin{equation}
    E(\phi) = \frac{1}{2} L I(\phi)^2 + E_J(\phi)
\end{equation}
composed out of inductive energy and junction potential energy.

\vspace{10 pt}
{\bf Finite momentum superconductivity from orbital effect versus spin-orbit and Zeeman effect:}
We discuss additional numerical calculations for the spectrum, ground state phase difference $\phi_0$ and diode efficiency $\eta$ varying the strength of spin-orbit coupling $\alpha$, the $g$-factor, and the distance $d$ between the two-dimensional electron gas (2DEG) and the superconducting lead (S) determining the phase gradient $2q = 2\pi d B_\parallel / \Phi_0$. Table \ref{tabS1} summarizes the varied parameters with references to the figures. 

\begin{table*}[]
    \centering
    \begin{tabular*}{\textwidth}{c @{\extracolsep{\fill}} ccccc}
    \hline \hline
         $\alpha$ & $g$ & $d$ & $w_s$ & spectrum & $\phi_0$, $\eta$ \\ \hline
         15 meV nm & -10 & 15 nm & 400 nm & Fig.~\ref{fig04} and \ref{figS1} & Fig.~\ref{fig04} and \ref{figS2} \\
         0 & 0 & 15 nm & 400 nm & Fig.~\ref{figS3} & Fig.~\ref{figS4} \\
         15 meV nm & -10 & 0 & 400 nm & Fig.~\ref{figS5} & Fig.~\ref{figS6} \\
         15 meV nm & -60 & 15 nm & 400 nm & Fig.~\ref{figS7} & Fig.~\ref{figS8} \\
         90 meV nm & -60 & 15 nm & 400 nm & Fig.~\ref{figS9} & Fig.~\ref{figS10} \\
         15 meV nm & -10 & 15 nm & 200 nm & Fig.~\ref{figS11} & Fig.~\ref{figS12} \\
         15 meV nm & -10 & 15 nm & 600 nm & Fig.~\ref{figS13} & Fig.~\ref{figS14} \\
         \hline \hline
    \end{tabular*}
    \caption{Parameters used in the calculations presented in the theory figures: spin-orbit strength $\alpha$, $g$-factor $g$, distance between the superconductor and the 2DEG $d$, and width of the superconducting lead $w_s$. The columns labeled 'spectrum' and '$\phi_0$, $\eta$' contain the references to the figures containing the calculations of the spectrum and ground state phase difference $\phi_0$ and diode efficiency $\eta$, respectively. All simulations use an effective mass $m^* = 0.026m_e$, chemical potential $\mu = 1.5$~meV, strength of the proximity induced pairing $|\Delta| = 0.2$~meV, and width $w_n = 100$~nm of the normal barrier.}
    \label{tabS1}
\end{table*}

For typical material and geometry parameters $\alpha = 15$~meV~nm, $g=-10$, and a distance $d = 15$~nm between the 2DEG and S as applied for the numerical calculations shown in Fig.~\ref{fig04} in the main text and Fig.~\ref{figS1} and Fig.~\ref{figS2} in the supplement, we reproduce qualitatively the spectral non-reciprocity of the experimental results. Turning off the spin-orbit coupling $\alpha$ and $g$-factor only has a weak effect on the spectrum and preserves the non-reciprocity, Fig.~\ref{figS3}. The calculated diode efficiency including spin-orbit coupling and Zeeman energy shown in Fig.~\ref{fig04} and \ref{figS2} exhibits a reversal of the diode asymmetry followed by a peak in diode efficiency at $B = 250$~mT. This feature is absent in the diode efficiency calculations without spin-orbit and Zeeman energy, indicating an origin in terms of the interplay of spin-orbit, Zeeman energy and the superconducting phase gradient due to the orbital effect \cite{YokoyamaPRB2014}. 
When the orbital effect of the magnetic field is neglected by setting the distance $d$ between 2DEG and S to zero, for typical parameters of spin-orbit coupling $\alpha = 15$~meV~nm and $g$-factor $g=-10$, the numerical calculations, Fig~\ref{figS5}, show an almost reciprocal spectrum. As a consequence, the ground state phase difference $\phi_0$ is almost zero and the diode efficiency $\eta$, Fig.~\ref{figS6}, is small for the investigated strength of the in-plane magnetic field $B_\parallel < 300$~mT. 

We note that for small magnetic fields $\hbar q v_F < |\Delta|$, with the Fermi velocity $v_F = \sqrt{2\mu/m^*} = 1.5\times 10^5 \ \text{m}/\text{s}$ and proximity induced pairing potential $|\Delta| = 0.2$~meV, the numerically calculated phase minimum $\phi_0$ as a function of $B_\parallel$ shown in Figs.~\ref{figS2} and \ref{fig04}(d), Fig.~\ref{figS4}, and Fig.~\ref{figS6} matches the analytical expression
\begin{equation}
    \phi_0 = - 2 q w_\text{n} + \text{arcsin} \left( \frac{2 q v_F}{\pi |\Delta|} \right)
    \label{eq:theory_phi0}
\end{equation}
which has been calculated for a one-dimensional wire with infinite superconducting leads $w_{\text{s}} \to \infty$ and without spin-orbit coupling or Zeeman effect \cite{DavydovaSci2022}. The first term in Eq.~\ref{eq:theory_phi0} is a phase bias due to the rotation of the phase of the superconducting order parameter as the Cooper pairs traverse the normal barrier, and the second term is a contribution from the continuum of states with energy above $\Delta$ due to the asymmetric deformation of the spectrum. In the presence of a large inductance, the second term shifts the phase-bias window selected by the inductance away from $\phi + 2 q w_n \mod 2 \pi = 0$ where the ABSs connect to the continuum \cite{DavydovaSci2022} and the ABS spectrum is most symmetric. Thereby, the visible asymmetry in the phase-bias window is enhanced. 

In the idealized one-dimensional wire, the junction gap closes at $\hbar q v_F = |\Delta|$. At large fields $\hbar q v_F \gg |\Delta|$, the continuum contribution in the analytical model saturates such that $\phi_0 = - 2 q w_\text{n} - \pi/2$ and the current phase relation approaches a $\pi/2$-shifted sinusoidal form. In our system, the junction gap closes at field strength larger than $\hbar q v_F = |\Delta|$, which we attribute mostly to the finite width $w_s$ of the superconducting leads. Numerical calculations increasing $w_s$ show increasing spectral non-reciprocity and the critical magnetic field strength to close the junction gap approach the theoretical limit $\hbar q v_F = |\Delta|$, see discussion below and Figs.~\ref{figS11} to ~\ref{figS14} for the numerical data.

We further investigate whether stronger spin-orbit coupling and $g$-factor can reproduce similar results. Increasing the $g$-factor to $g=-60$ while keeping the spin-orbit strength at $\alpha = 15$~meV~nm, the calculated spectrum shown in Fig,~\ref{figS7} remains approximately symmetric around $\phi = 0$ and $\pi$ superconducting phase bias. At $B_\parallel = 120$~mT, the junction undergoes a zero-$\pi$-transition (see Fig.~\ref{figS8}) coinciding with a closure of the gap in the junction at zero phase bias $\phi = 0$. This zero-$\pi$-transition is driven by the Zeeman energy $E_Z = g \mu_B B_\parallel / 2$ shifting one spin band across zero energy when $E_Z = |\Delta|$ at $|B_\parallel| = 2 |\Delta| / |g| \mu_B \sim 120$~mT, analogous to a zero-$\pi$-transition in superconductors coupled by a quantum dot occupied by an odd number of electrons \cite{BulaevskiiJETP1977, vanDamNature2006, WhiticarPRB2021}. Despite this band inversion, also a phase difference $\theta_B = |g \mu_B B_\parallel| w_n /\hbar v_F$ can give rise to 0-$\pi$-transitions when $\theta_B \mod \pi = \pi/2$ \cite{YokoyamaPRB2014}, however this is not the mechanism in our case since $\theta_B \ll \pi$ for the investigated magnetic field strength $B_\parallel < 300$~mT.

The presence of inversion-symmetry breaking spin-orbit coupling in the clean system gives rise to the two peaks in the diode efficiency with reversal of the diode asymmetry at the 0-$\pi$ transition as shown in Fig.~\ref{figS8}, in agreement with results from Ref.~\cite{YokoyamaPRB2014}.

For enhanced spin-orbit coupling $\alpha = 90$~meV~nm and $g$-factor $g=-60$, the model spectrum achieves nonreciprocal values comparable to experiment, see Fig.~\ref{figS9}. This non-reciprocity gives rise to a shift of ground state minimum (Fig.~\ref{figS10}), which agrees well with the prediction from Ref.~\onlinecite{BuzdinPRL2008} in terms of the rotation of the proximity induced order parameter in the normal barrier
\begin{equation}
    \phi_0^\text{rot} = \frac{2 \alpha^2 g \mu_B B_\parallel}{\hbar^2 \bar{v}_y^2}
    \label{eq:S_theory_phi0_phase_rotation}
\end{equation}
where we use the inversion averaged Fermi velocity of channels propagating along the junction direction $y$ \cite{YokoyamaPRB2014}, $\bar{v}_y^{(-1)} = \frac{1}{v_F} \int_0^{k_F} dk_x 1/v_y(k_x) = \frac{\pi}{2} v_F^{-1}$ where $v_y(k_x) = \sqrt{v_F^2 - \hbar^2 k_x^2/m^2}$ and $v_F = \sqrt{2\mu/m + \alpha^2/\hbar^2}$ the Fermi velocity in the presence of spin-orbit coupling $\alpha$. The formula yields the same result for both spin-orbit split Fermi surfaces. The phase minimum jumps by a small amount when the Zeeman energy is equal to the proximity induced pairing, similar to the 0-$\pi$ transition for weaker spin-orbit coupling. We note that in helical channels with Zeeman field pointing in the same direction as the spin-momentum locking, the Zeeman field can be transformed into a superconducting phase gradient by a spatial gauge transformation \cite{DolciniPRB2015}, yielding an analogous diode effect as obtained from finite momentum pairing in Ref.~\cite{DavydovaSci2022}.

In conclusion, we believe that the dominant contribution to the non-reciprocity of spectrum and supercurrents in our device is the finite Cooper pair momentum due to the orbital effect of the in-plane magnetic field $B_\parallel$. An explanation in terms of spin-orbit and Zeeman coupling without the orbital effect (setting $d = 0$) requires unrealistically large spin-orbit strength $\alpha = 90$~meV~nm and $g$-factor $g = -60$.

\vspace{10pt}
{\bf Width of the superconducting leads:}
Figures \ref{figS11} to \ref{figS14} show the calculated spectrum and extracted phase minimum $\phi_0$ and diode efficiency $\eta$ for a system with typical parameters $\alpha = 15$~meV~nm, $g = -10$, for different width of the superconducting leads $w_s = 200$~nm and $w_s = 600$~nm. The calculated spectra in Fig.~\ref{figS11}, Fig.~\ref{fig04} and Fig.~\ref{figS1}, and Fig.~\ref{figS13} for $w_s = 200$~nm, 400~nm, and 600~nm, respectively, indicate that the visible non-reciprocity increases with the width $w_s$. Also the phase minimum $\phi_0$ and diode efficiency $\eta$ quantifying the non-reciprocity shown in Fig.~\ref{figS12}, Fig.~\ref{fig04} and Fig.~\ref{figS2}, and Fig.~\ref{figS14} for $w_s = 200$~nm, 400~nm, and 600~nm, respectively, increase with the width $w_s$. In the limit of $w_s \to \infty$, we expect the non-reciprocity to saturate and reproduce approximately the analytical results from Ref.~\onlinecite{DavydovaSci2022}, up to corrections from spin-orbit coupling, Zeeman effect and the larger number of channels in our junction. We expect the length scale above which the $w_s \to \infty$ regime is reached to be the coherence length in the proximitized two-dimensional electron gas $\xi_\text{p} = \hbar v_F / \pi \Delta \sim 150$~nm. This length scale governs the distance quasiparticles with energy $E \ll |\Delta|$ travel within the proximitized regions before they enter the superconducting leads and Andreev reflect as a hole. 

\vspace{10pt}
{\bf Supplementary experimental data:}
Fig.~\ref{figS15} shows the Andreev bound state spectrum in a transverse in-plane magnetic field $B_t$, applied perpendicular to the S-N interfaces. The proximity induced gap closes at around $B_t = 120$~mT while no significant phase-asymmetry is observed.
This observation is consistent with an expectation based on the symmetry of the device in this configuration. Neglecting disorder, this configuration preserves a mirror symmetry $y \to - y$ exchanging the superconducting leads such that the current phase relation and spectrum should be symmetric. This also holds in the presence of spin-orbit coupling \cite{BaumgartnerNNano2022}. Intuitively, the Cooper pair momentum induced by the orbital coupling of $B_t$ is perpendicular to the supercurrent direction such that the Doppler shift of the Bogoliubov quasiparticles does not lead to a non-reciprocity in spectrum and current-phase relation. 

Fig.~\ref{figS16} and \ref{figS17} display the differential conductance matrix in devices 2 and 3 at zero in-plane magnetic field and at $B_\parallel = 150$~mT and $180$~mT, respectively. These measurements are in agreement with the results presented in the main text.

\clearpage

\begin{figure*}
\begin{tabular}{ll}
  (a)   & (b) \\
  \includegraphics{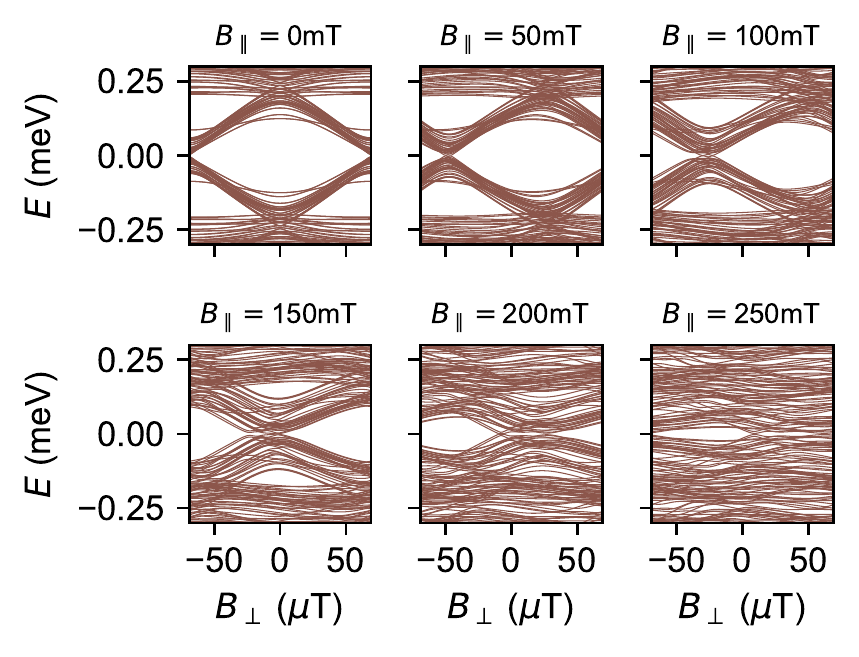}
  & 
  \includegraphics{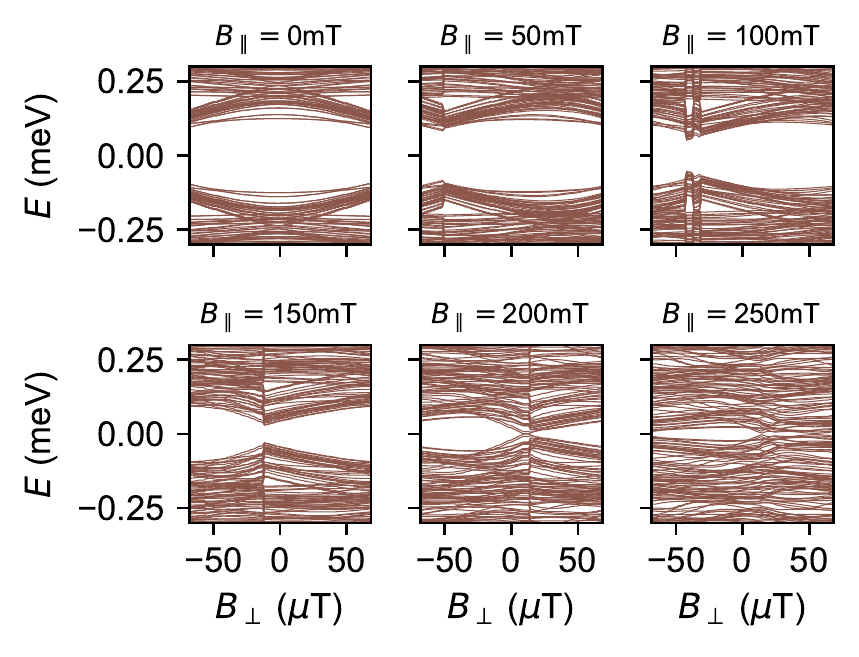}
\end{tabular}
\caption{
\textbf{Theory.} Calculated spectrum (a) without inductance $L = 0$ and (b) with inductance $L = 1nH$. Here we use the same parameters as in Fig.~\ref{fig04} the main text, a center-to-center distance between 2DEG and S $d=15$~nm, a g-factor $g=-10$ and spin-orbit coupling strength $\alpha=15$~nm~meV. The perpendicular magnetic field range $B_\perp \in [-\Phi_0 / A_\text{loop}, \Phi_0 / A_\text{loop}]$ corresponds to a superconducting phase bias $\phi \in [-\pi, \pi]$.
}
\label{figS1}
\end{figure*}

\begin{figure}
    \centering
    \includegraphics{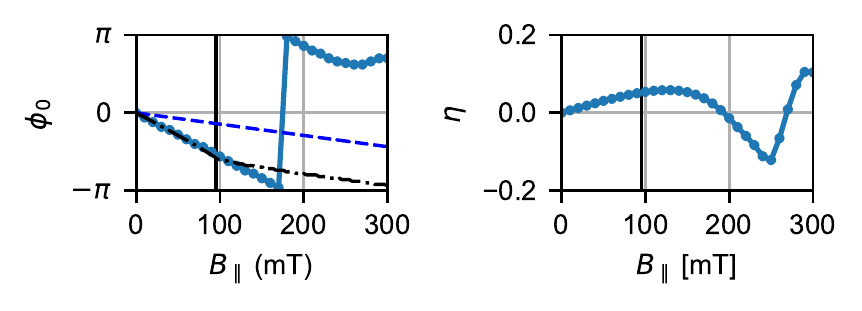}
    \caption{\textbf{Theory.} (a) Calculated phase difference $\phi_0$ that minimize the ground state energy (full line), phase-bias from the phase gradient $- 2 q w_n$ (blue dashed line), and analytical phase minimum $\phi_0$ including the continuum contribution given by Eq.~\eqref{eq:theory_phi0} (black dashed-dotted line). The vertical line denotes the magnetic field strength $\hbar v_F q = \Delta$. (b) Calculated diode efficiency $\eta$. Here we use the same parameters as in Fig.~\ref{figS1} and Fig.~\ref{fig04} in the main text.}
    \label{figS2}
\end{figure}

\begin{figure*}
\begin{tabular}{ll}
  (a)   & (b) \\
  \includegraphics{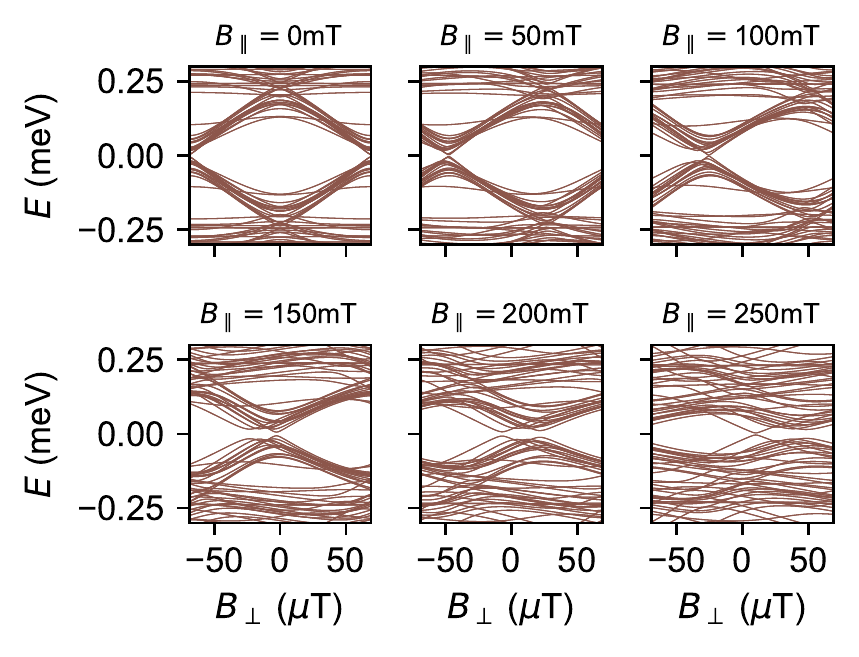}
  & 
  \includegraphics{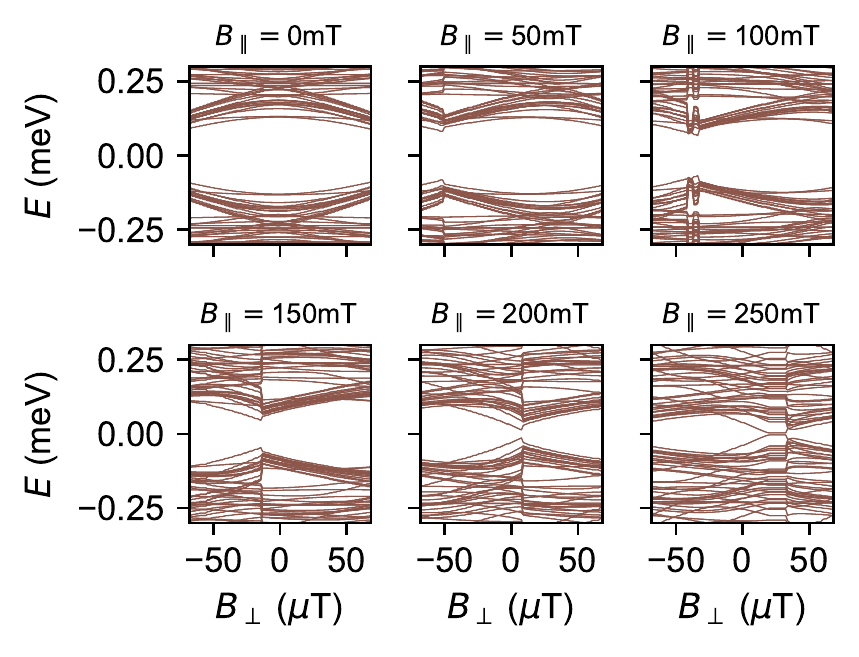}
\end{tabular}
\caption{
\textbf{Theory.} Calculated spectrum (a) without inductance $L = 0$ and (b) with inductance $L = 1nH$. Here we neglect Zeeman energy and spin-orbit coupling, $g = 0$ and $\alpha = 0$. The distance between the centers of 2DEG and S is $d = 15$~nm, such that the orbital effect is included with phase gradient $q = 2\pi B_\parallel d / \Phi_0$.
}
\label{figS3}
\end{figure*}

\begin{figure}
    \centering
    \includegraphics{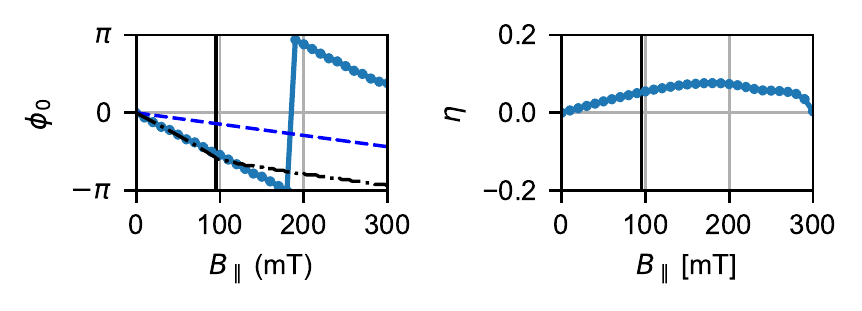}
    \caption{\textbf{Theory.} (a) Calculated phase difference $\phi_0$ that minimize the ground state energy (full line), phase-bias from the phase gradient $- 2 q w_n$ (blue dashed line), and analytical phase minimum $\phi_0$ including the continuum contribution given by Eq.~\eqref{eq:theory_phi0} (black dashed-dotted line). The vertical line denotes the magnetic field strength $\hbar v_F q = \Delta$. (b) Calculated diode efficiency $\eta$. Here we use the same parameters as in Fig.~\ref{figS2}: We neglect Zeeman energy and spin-orbit coupling, $g = 0$ and $\alpha = 0$. The distance between the centers of 2DEG and S is $d = 15$~nm, such that the orbital effect is included with phase gradient $q = 2\pi B_\parallel d / \Phi_0$.}
    \label{figS4}
\end{figure}

\begin{figure*}
\begin{tabular}{ll}
  (a)   & (b) \\
  \includegraphics{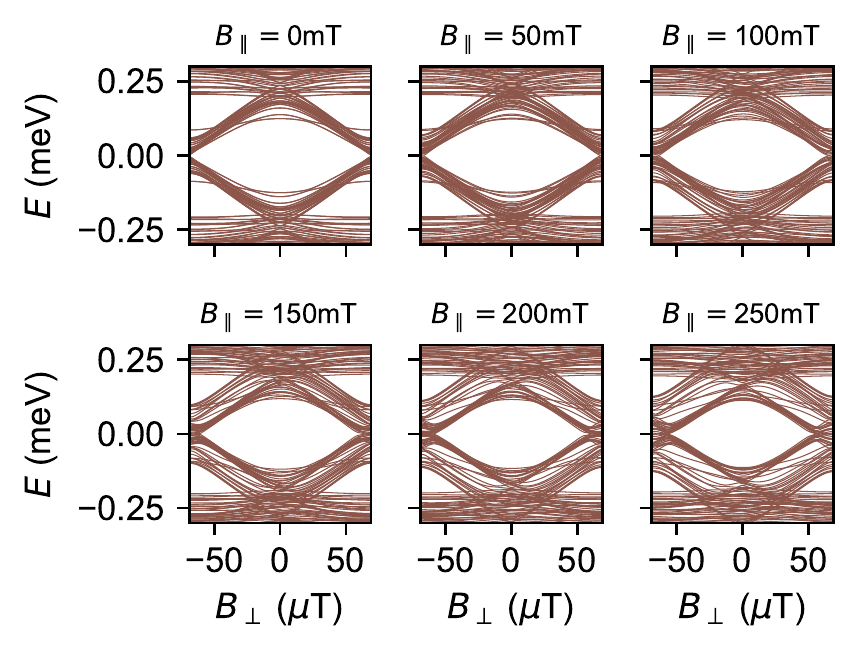}
  & 
  \includegraphics{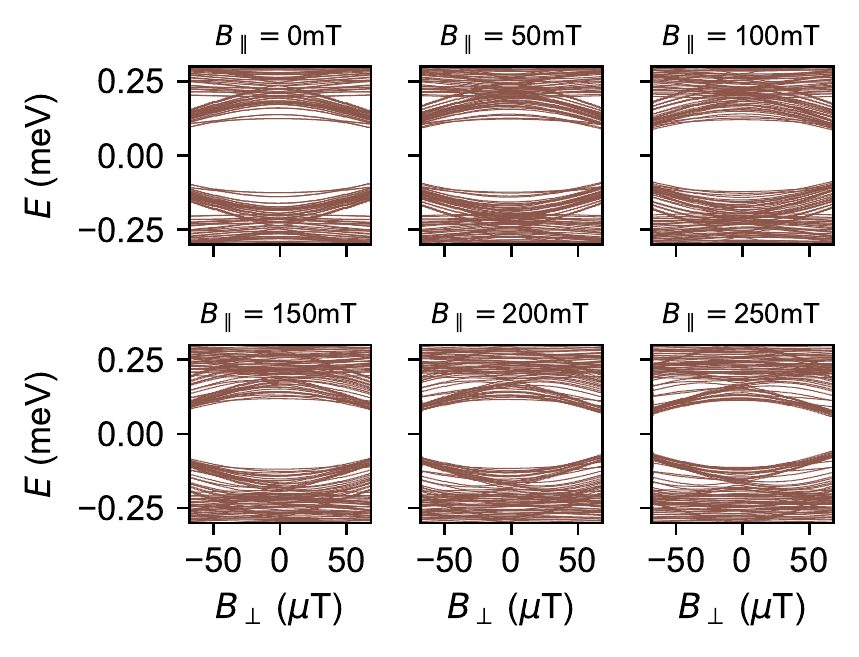}
\end{tabular}
\caption{
\textbf{Theory.} Calculated spectrum (a) without inductance $L = 0$ and (b) with inductance $L = 1nH$. Here we neglect the orbital effect of the magnetic field by setting the distance $d$ between the centers of 2DEG and S to zero, such that the proximity-induced phase gradient $q$ vanishes. For Zeeman-energy and spin-orbit coupling, we use $g = -10$ and $\alpha = 15$~nm~meV.
}
\label{figS5}
\end{figure*}

\begin{figure}
    \centering
    \includegraphics{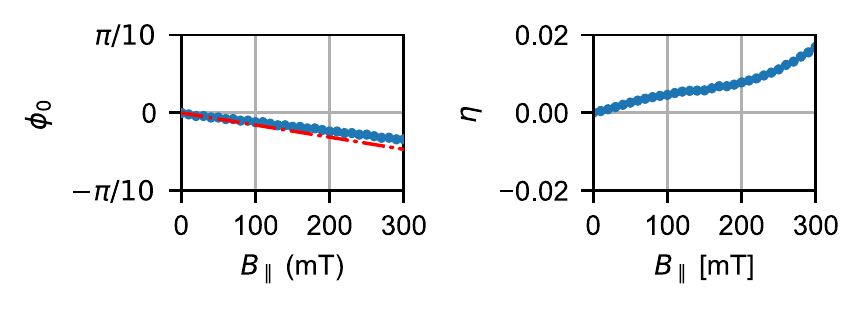}
    \caption{\textbf{Theory.} (a) Calculated phase difference $\phi_0$ that minimize the ground state energy (full line). The red dashed-dotted line denotes the phase difference from the superconducting phase rotation due to spin-orbit and Zeeman coupling as given by Eq.~\eqref{eq:S_theory_phi0_phase_rotation}. (b) Calculated diode efficiency $\eta$. Here we use the same parameters as in Fig.~\ref{figS5}.}
    \label{figS6}
\end{figure}

\begin{figure*}
\begin{tabular}{ll}
  (a)   & (b) \\
  \includegraphics{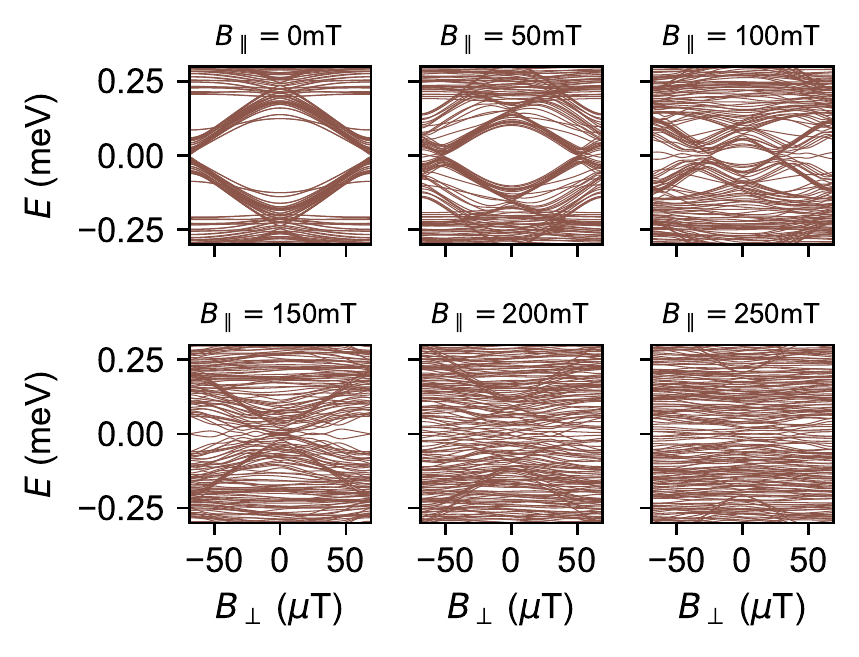}
  & 
  \includegraphics{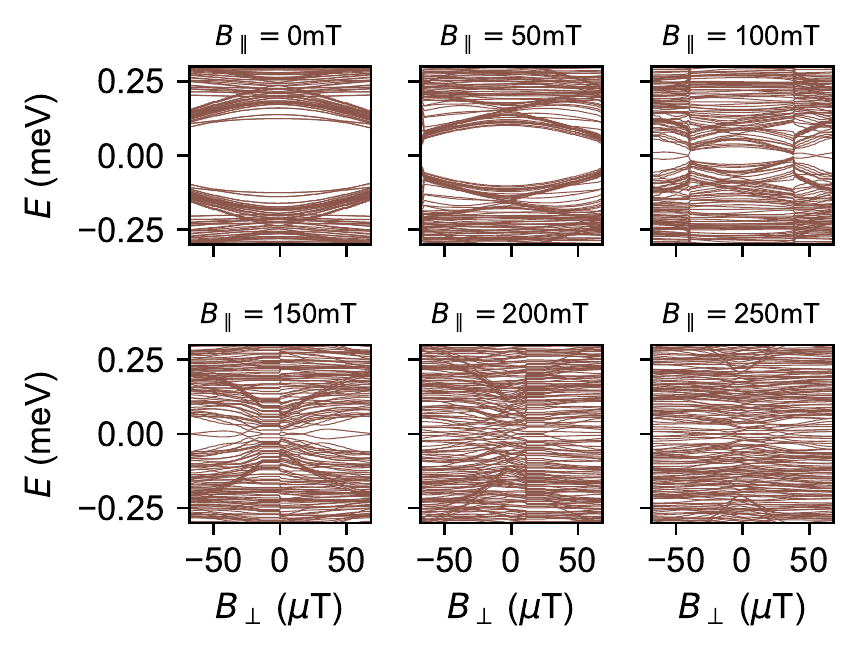}
\end{tabular}
\caption{
\textbf{Theory.} Calculated spectrum (a) without inductance $L = 0$ and (b) with inductance $L = 1$~nH. Here we neglect the orbital effect of the magnetic field by setting the distance $d$ between the centers of 2DEG and S to zero, such that the proximity-induced phase gradient $q$ vanishes. For Zeeman-energy and spin-orbit coupling, we use $g = -60$ and $\alpha = 15$~nm~meV.
}
\label{figS7}
\end{figure*}

\begin{figure}
    \centering
    \includegraphics{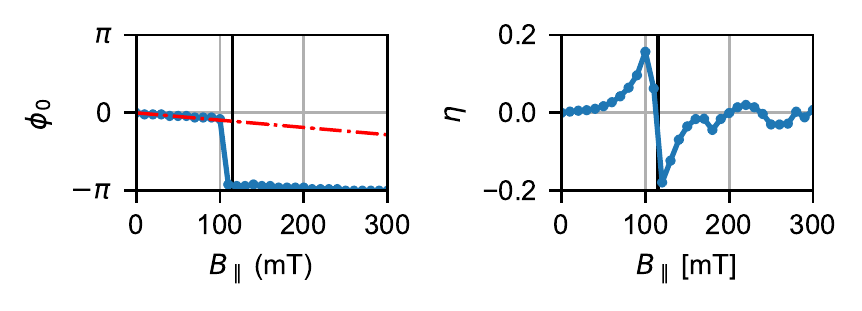}
    \caption{\textbf{Theory.} (a) Calculated phase difference $\phi_0$ that minimize the ground state energy (full line). The red dashed-dotted line denotes the phase difference from the superconducting phase rotation due to spin-orbit and Zeeman coupling as given by Eq.~\eqref{eq:S_theory_phi0_phase_rotation}. The vertical line denotes the magnetic field strength at which the Zeeman field is equal to the pairing strength $g \mu_B B_\parallel / 2 = |\Delta|$. (b) Calculated diode efficiency $\eta$. Here we use the same parameters as in Fig.~\ref{figS7}.}
    \label{figS8}
\end{figure}

\begin{figure*}
\begin{tabular}{ll}
  (a)   & (b) \\
  \includegraphics{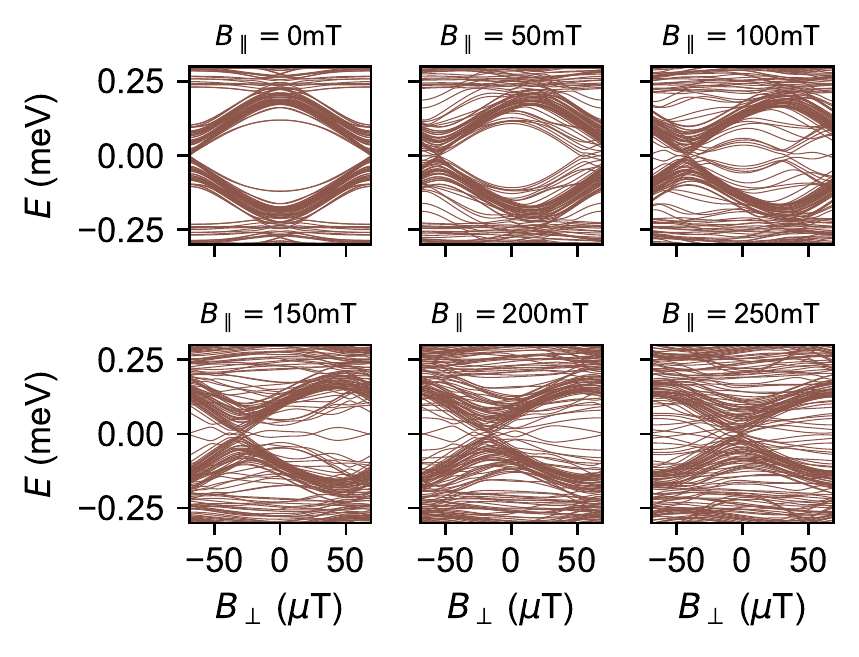}
  & 
  \includegraphics{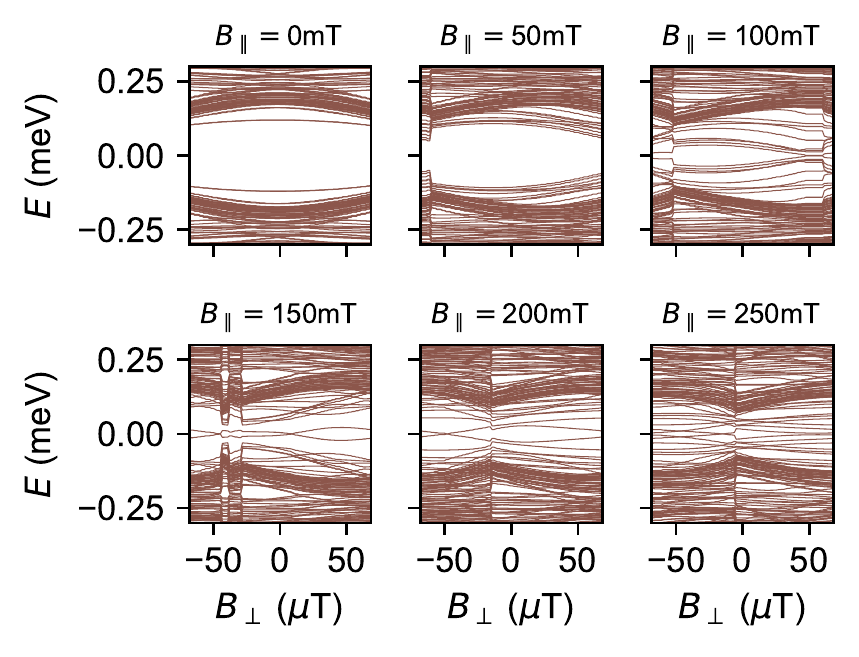}
\end{tabular}
\caption{
\textbf{Theory.} Calculated spectrum (a) without inductance $L = 0$ and (b) with inductance $L = 1$~nH. Here we neglect the orbital effect of the magnetic field by setting the distance $d$ between the centers of 2DEG and S to zero, such that the proximity-induced phase gradient $q$ vanishes. For Zeeman-energy and spin-orbit coupling, we use $g = -60$ and $\alpha = 90$~nm~meV.
}
\label{figS9}
\end{figure*}

\begin{figure}
    \centering
    \includegraphics{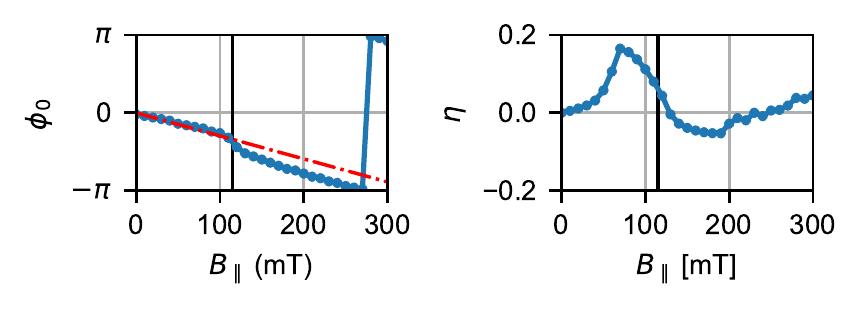}
    \caption{\textbf{Theory.} (a) Calculated phase difference $\phi_0$ that minimize the ground state energy (full line). The red dashed-dotted line denotes the phase difference from the superconducting phase rotation due to spin-orbit and Zeeman coupling as given by Eq.~\eqref{eq:S_theory_phi0_phase_rotation}. The vertical line denotes the magnetic field strength at which the Zeeman field is equal to the pairing strength $g \mu_B B_\parallel / 2 = |\Delta|$. (b) Calculated diode efficiency $\eta$. Here we use the same parameters as in Fig.~\ref{figS9}.}
    \label{figS10}
\end{figure}

\begin{figure*}
\begin{tabular}{ll}
  (a)   & (b) \\
  \includegraphics{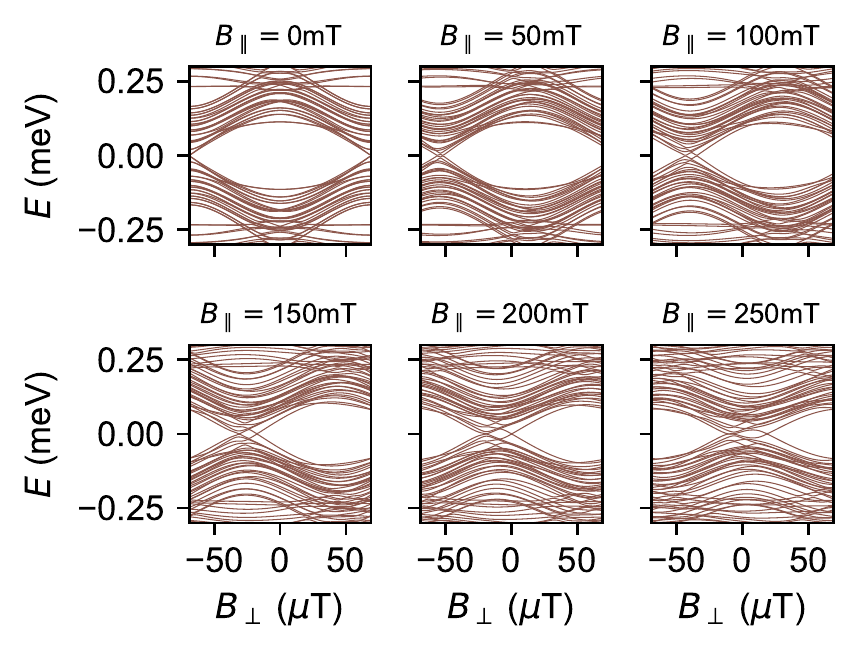}
  & 
  \includegraphics{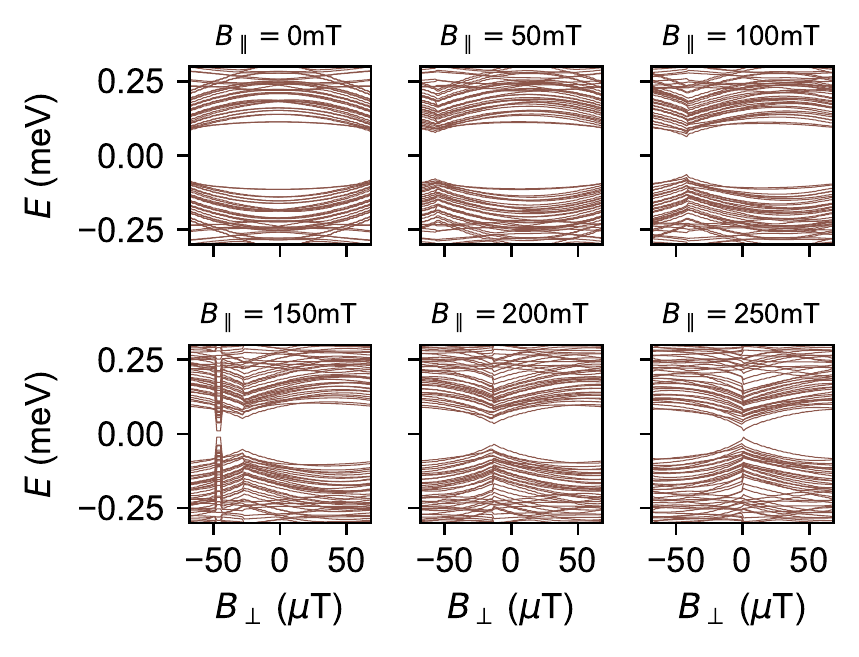}
\end{tabular}
\caption{
\textbf{Theory.} Calculated spectrum (a) without inductance $L = 0$ and (b) with inductance $L = 1$~nH. Here we use the same material parameter as in the main text: a center-to-center distance between 2DEG and S $d=15$~nm, a g-factor $g=-10$ and spin-orbit coupling strength $\alpha=15$~nm~meV. Here, the width of the superconducting bands is reduced to $w_\text{S} = 200$~nm.
}
\label{figS11}
\end{figure*}

\begin{figure}
    \centering
    \includegraphics{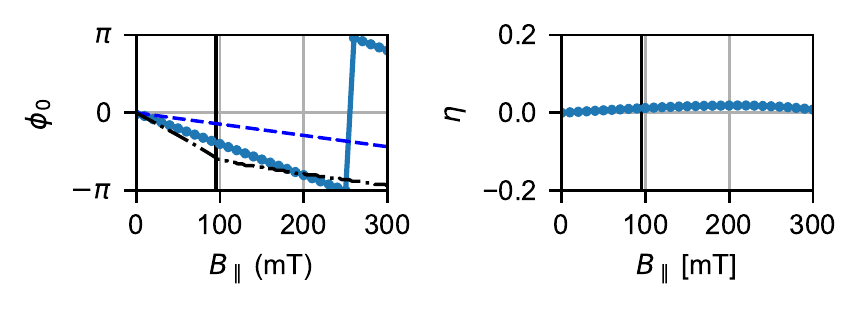}
    \caption{\textbf{Theory.} (a) Calculated phase difference $\phi_0$ that minimize the ground state energy (full line), phase-bias from the phase gradient $- 2 q w_n$ (blue dashed line), and analytical phase minimum $\phi_0$ including the continuum contribution given by Eq.~\eqref{eq:theory_phi0} (black dashed-dotted line). The vertical line denotes the magnetic field strength $\hbar v_F q = \Delta$. and (b) calculated diode efficiency $\eta$. Here we use the same parameters as in Fig.~\ref{figS11}.}
    \label{figS12}
\end{figure}

\begin{figure*}
\begin{tabular}{ll}
  (a)   & (b) \\
  \includegraphics{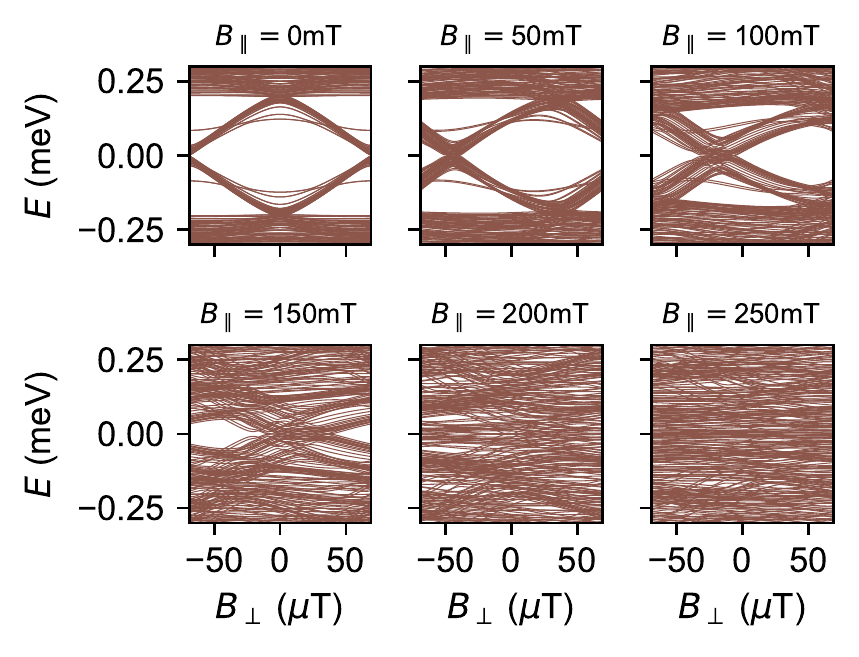}
  & 
  \includegraphics{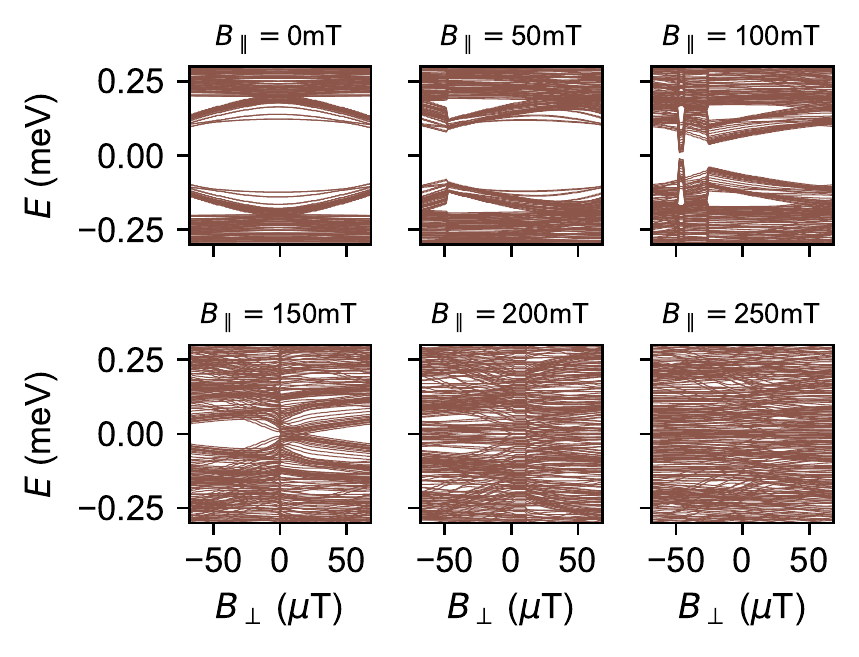}
\end{tabular}
\caption{
\textbf{Theory.} Calculated spectrum (a) without inductance $L = 0$ and (b) with inductance $L = 1$~nH. Here we use the same material parameter as in the main text: a center-to-center distance between 2DEG and S $d=15$~nm, a g-factor $g=-10$ and spin-orbit coupling strength $\alpha=15$~nm~meV. Here, the width of the superconducting bands is increased to $w_\text{S} = 600$~nm.
}
\label{figS13}
\end{figure*}

\begin{figure}
    \centering
    \includegraphics{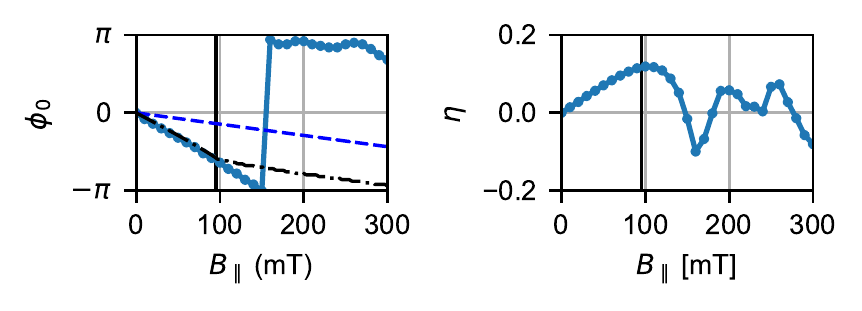}
    \caption{\textbf{Theory.} (a) Calculated phase difference $\phi_0$ that minimize the ground state energy (full line), phase-bias from the phase gradient $- 2 q w_n$ (blue dashed line), and analytical phase minimum $\phi_0$ including the continuum contribution given by Eq.~\eqref{eq:theory_phi0} (black dashed-dotted line). The vertical line denotes the magnetic field strength $\hbar v_F q = \Delta$. and (b) calculated diode efficiency $\eta$. Here we use the same parameters as in Fig.~\ref{figS14}.}
    \label{figS14}
\end{figure}

\begin{figure*}[t]
\includegraphics[width=1.0\textwidth]{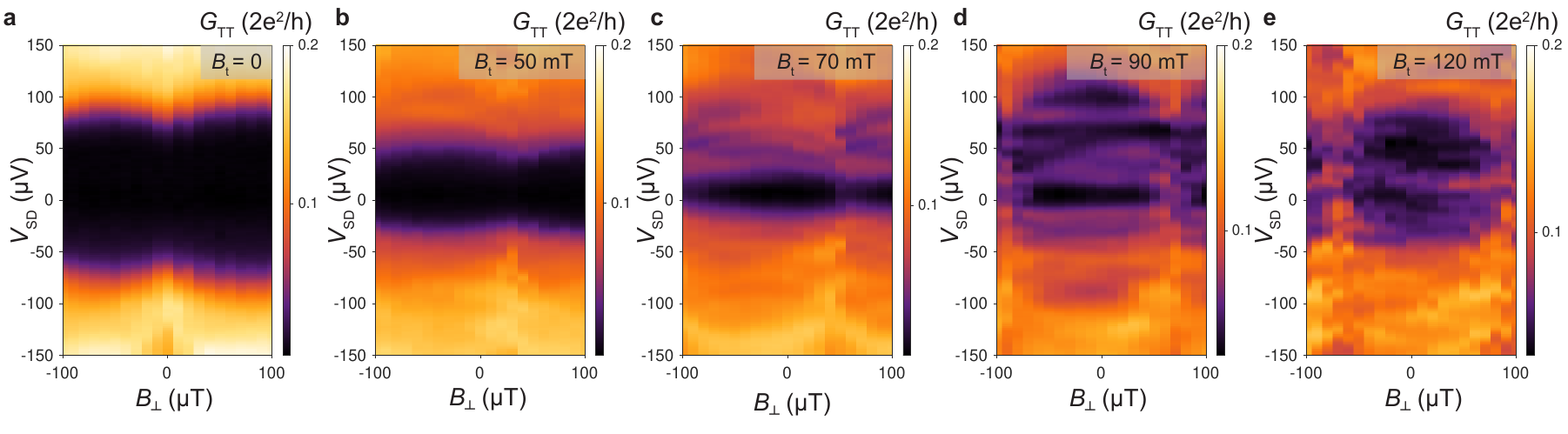}
\caption{\label{figS15} 
{\bf Device 1: Andreev bound state spectrum in transverse magnetic field.} (a)--(e) Differential conductance measured in Device 1 as a function of $B_\perp$ at different values of a transverse magnetic field $B_t$, applied in the plane of the sample and perpendicular to the S-N interfaces. The ABS spectrum is roughly phase-symmetric within each flux lobe for all values of $B_t$. The spectral gap closes at $B_t \simeq 120$~mT.}
\end{figure*}

\begin{figure*}[t]
\includegraphics[width=1\textwidth]{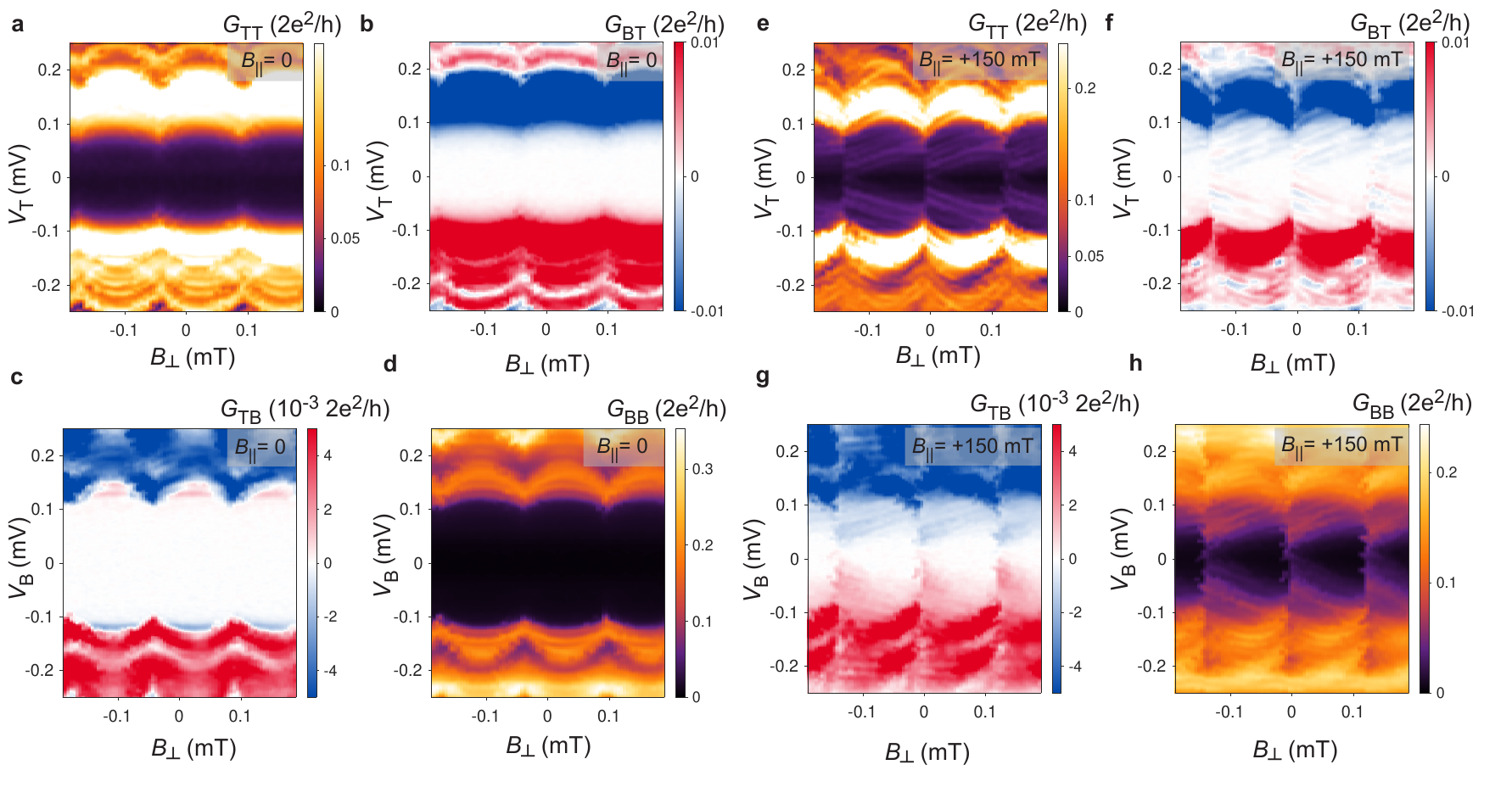}
\caption{\label{figS16} {\bf Device 2: Nonreciprocal differential conductance matrix.} Local conductances (a) $G_{\rm TT}$ and (d) $G_{\rm BB}$ , and nonlocal conductances (b) $G_{\rm BT}$ and (c) $G_{\rm TB}$, measured as a function of source-drain bias $V_{\rm SD}$ and out-of-plane magnetic field $B_\perp$ at $B_\parallel=0$, in Device 2. (e)-(h) Differential conductance matrix measured at $B_\parallel=150$~mT.}
\end{figure*}

\begin{figure*}[t]
\includegraphics[width=1\textwidth]{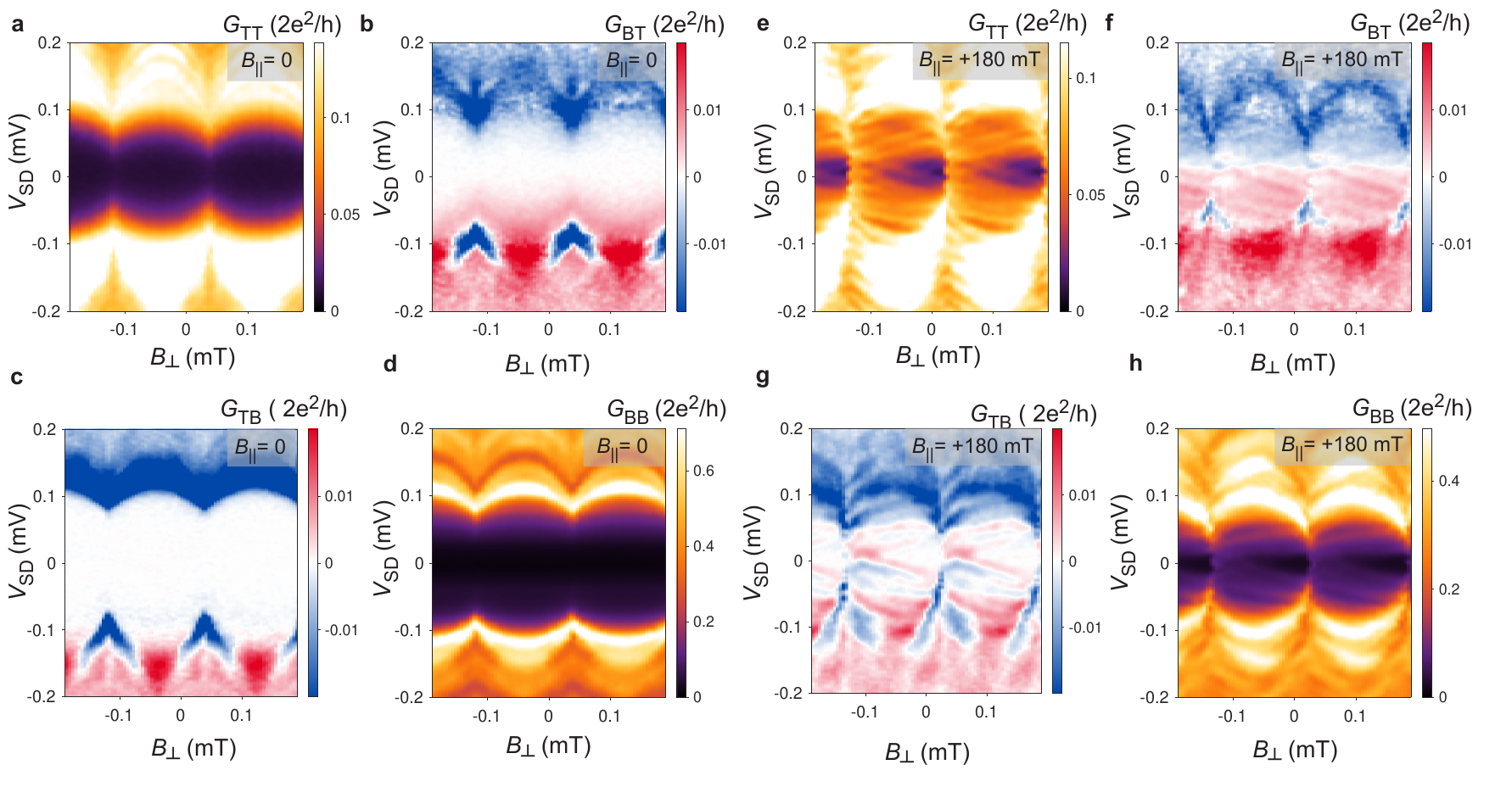}
\caption{\label{figS17} {\bf Device 3: Nonreciprocal differential conductance matrix.} Local conductances (a) $G_{\rm TT}$ and (d) $G_{\rm BB}$ , and nonlocal conductances (b) $G_{\rm BT}$ and (c) $G_{\rm TB}$, measured as a function of source-drain bias $V_{\rm SD}$ and out-of-plane magnetic field $B_\perp$ at $B_\parallel=0$, in Device 3. (e)-(h) Differential conductance matrix measured at $B_\parallel=180$~mT. Both local and nonlocal conductances show phase-asymmetric Andreev bound states that have a smaller gap at the left end of a flux lobe and a larger gap at the right end.}
\end{figure*}

\end{document}